\documentclass[fleqn,usenatbib]{mnras}

\usepackage{newtxtext,newtxmath}
\usepackage[T1]{fontenc}
\usepackage{soul}

\let\VANthebibliography\thebibliography
\def\thebibliography{\DeclareRobustCommand{\VAN}[3]{##3}\VANthebibliography}

\usepackage{graphicx}	
\usepackage{amsmath}	
\usepackage{appendix}
\usepackage{lscape}
\usepackage{ulem}
\usepackage[detect-none]{siunitx}
\newcommand{\emerlin}{\textit{e}{-MERLIN}}
\newcommand{\emerge}{\textit{e}{-MERGE}}
\newcommand{\microJybm}{\,\mathrm{\umu Jy\,beam^{-1}}}

\title[VLBA CANDELS GOODS-North Survey \-- II]{The VLBA CANDELS GOODS-North Survey. II \-- Wide-field source catalogue comparison between the VLBA, EVN, \textit{e}-MERLIN and VLA }

\author[Njeri A. et al.]
{Ann ~Njeri,$^{1,2}$\thanks{E-mail: ann.njeri@newcastle.ac.uk}
Roger.~P.~Deane,$^{3,4}$
J.~F.~Radcliffe,$^{2,4}$
R.~J.~Beswick,$^{2}$
A.~P.~Thomson,$^{2}$
T.~W.~B.~Muxlow,$^{2}$
\newauthor
M.~A.~Garrett,$^{2}$
C.~M.~Harrison$^{1}$
\\
$^{1}$School of Mathematics, Statistics \& Physics, Newcastle University, NE1 7RU, Newcastle Upon Tyne, UK\\
$^{2}$Jodrell  Bank  Centre  for  Astrophysics,  School  of  Physics  \&  Astronomy,  The  University  of  Manchester,  Alan  Turing  Building, Oxford Road, Manchester M13 9PL, UK\\
$^{3}$Wits Centre for Astrophysics, University of the Witwatersrand, Private Bag 3, 2050, Johannesburg, South Africa\\
$^{4}$Department of Physics, University of Pretoria, Hatfield, Pretoria, 0028, South Africa\\
}

\date{Accepted 2024 February 01. Received 2024 January 31; in original form 2023 June 15}

\pubyear{2024}

\begin{document}
\label{firstpage}
\pagerange{\pageref{firstpage}--\pageref{lastpage}}
\maketitle

\begin{abstract}
Deep radio surveys of extragalactic legacy fields trace a large range of spatial and brightness temperature sensitivity scales, and therefore have differing biases to radio-emitting physical components within galaxies. This is particularly true of radio surveys performed at $\lesssim 1\,\mathrm{arcsec}$ angular resolutions, and so robust comparisons are necessary to better understand the biases present in each survey. We present a multi-resolution and multi-wavelength analysis of the sources detected in a new Very Long Baseline Array (VLBA) survey of the CANDELS GOODS-North field. For the 24 VLBA-selected sources described in Paper I, we augment the VLBA data with EVN data, and $\sim0.1\mbox{--}1$\,arcsecond angular resolution data provided by VLA and \emerlin{}. This sample includes new AGN detected in this field, thanks to a new source extraction technique that adopts priors from ancillary multi-wavelength data. The high brightness temperatures of these sources ($T_{B}\gtrsim10^{6}$\,K) confirm AGN cores, that would often be missed or ambiguous in lower-resolution radio data of the same sources. Furthermore, only 15 sources are identified as `radiative' AGN based on available X-ray and infrared constraints. By combining VLA and VLBA measurements, we find evidence that the majority of the extended radio emission is also AGN dominated, with only 3 sources with evidence for extended potentially star-formation dominated radio emission. We demonstrate the importance of wide-field multi-resolution (arcsecond--milliarcsecond) coverage of the faint radio source population, for a complete picture of the multi-scale processes within these galaxies.
\end{abstract}

\begin{keywords}

methods: observational – techniques: high angular resolution – techniques: interferometric – galaxies: active –
galaxies: high-redshift – quasars: supermassive black holes.
\end{keywords}



\section{Introduction}

Understanding the evolution of star-formation and accretion through cosmic time, from the earliest stars, galaxies and black holes in the Universe to the present day, is one of the most important goals in astrophysics \citep[e.g.][]{Lilly1996, Madau2014}. This requires an understanding of the sequence of events from those distant epochs to the present and successfully resolving the processes involved, namely star formation and accretion. Therefore, it is necessary to disentangle and distinguish emission from these two processes \citep[e.g.][]{smolcic2009,Padovani2016}. Optical, infra-red, and X-ray, observations can be affected by dust obscuration and are an incomplete probes of these two processes. Radio observations on the other hand, are free from dust obscuration and therefore offer unbiased tracers for both star formation and accretion processes \citep[e.g.][]{Panessa2019}. 

Deep interferometric extragalactic radio continuum surveys such as the VLA-COSMOS \citep{Schinnerer2004,Smolcic2017}, MIGHTEE \citep{jarvis2017meerkat}, \textit{e}-MERGE \citep{Muxlow2020}, ATLAS \citep{Norris2006}, Lockman Hole \citep[e.g.,][]{de_Ruiter1997,2022sweijen}, the LOFAR Two-metre Sky Survey \citep[LoTSS;][]{2022shimwell}
and EMU \citep{Norris2011,2021norris} are prominent examples of studies of star formation and black hole accretion processes at high redshifts. These surveys have been facilitated by instruments such as the Karl G. Jansky Very Large Array (VLA), the enhanced-Multi Element Remotely Linked Interferometry Network (\emerlin{}), MeerKAT, the LOw Frequency ARray (LOFAR), the Murchison Widefield Array, and the Australian Square Kilometre Array Pathfinder (ASKAP).

 To date, these deep-field surveys have revealed both star-forming galaxies and Active Galactic Nuclei (AGN) populations over a wide range of luminosities and redshifts \citep[e.g.][]{Garrett2000,Muxlow2005,Padovani2009,Smolcic2017,Shimwell2019}. Multi-wavelength analysis reveal that star-forming galaxies begin to dominate at flux densities below $0.2\,$mJy at 1.4\,GHz \citep[e.g.][]{Morrison2010,Smolcic2017}. At this level, some galaxies show signatures of AGN activity at non-radio wavelengths (X-ray, mid-IR, and optical), but without signs of large-scale radio jets, and much weaker radio emission than the radio-loud AGN sources \citep[e.g.,][]{Kellermann2016,2021rad}.

Deep wide-field radio surveys of extragalactic legacy fields trace a large range of spatial scales and brightness temperature sensitivity, with differing sensitivities to the different radio-emitting components in galaxies at cosmologically-significant distances. This potentially opens an avenue for separating AGN and star formation contributions without the need for multi-wavelength data and its associated biases. At $\sim$GHz frequencies, the arcsecond angular resolution scales offer a view of the diffuse emission largely associated with star formation regions \citep[e.g.][]{Condon2012,Vernstrom2015}. Sub-arcsecond angular resolutions are sensitive to, and able to spatially resolve, emissions from both AGN and star formation \citep[e.g.,][]{Muxlow2020}, while milliarcsecond angular resolution scales are mostly sensitive to compact emission mainly associated with the cores of AGN \citep[e.g.][]{HerreraRuiz2017,Radcliffe2018,Njeri2023}. It is therefore crucial to better understand the resolution at which compact emission becomes more dominant than the diffuse emission and the associated radio properties. To achieve this, we also have to understand the biases present in each interferometric survey, and be able to reconcile them with one another with regards to; firstly, their observed astrophysical properties, such as their astrometric positions, morphology, flux density, variability, sizes, luminosity and brightness temperature, and, secondly, the instrumental and technical biases that arise from the differing survey designs, instruments and analysis techniques used. 


The Great Observatories Origins Deep Survey-North, \citep[GOODS-North;][]{Giavalisco2004}, is one of the best studied extragalactic legacy fields across the electromagnetic spectrum, but particularly so at radio wavelengths. This includes surveys spanning more than two decades using multiple instruments including the VLA \citep{Richards2000,Morrison2010,Murpy2017,Owen2018,Radcliffe2019,Gim19}, the Multi-Element Radio-Linked Interferometer Network \citep[MERLIN/\emerlin{};][]{Muxlow2005,Guidetti2013,Guidetti2017,Muxlow2020} and VLBI surveys with the European VLBI Network \citep[EVN;][]{Garrett2001,Chi2013,Radcliffe2018}. At 1.4 GHz, these surveys cover seven orders of magnitude in brightness temperature ($T_\mathrm{B}$$\sim$$10^{2}\mbox{--}10^{9}\,\mathrm{K}$), a wide range of sensitivity and an arcsec-scale to milliarcsec range of resolutions ($\theta_\mathrm{res}\sim0\farcs005\mbox{--}1\farcs5$). Combining such a range of $T_\mathrm{B}$ and $\theta_\mathrm{res}$ provides a powerful tool for probing and separating faint AGN from star formation and their symbiotic occurrences in this and other extragalactic fields \citep[e.g.][]{Chi2013,HerreraRuiz2017,Radcliffe2018,Njeri2023}.

We have extended the radio coverage in the Cosmic Assembly Near-IR Deep Extragalactic Legacy Survey (CANDELS) GOODS-North field, with a Very Long Baseline Array (VLBA) survey that covers the entirety of the $\sim$ 160 arcmin$^2$ area with quasi-uniform sensitivity (Deane et al., hereafter Paper I). This is a unique approach, which is in contrast to other wide-field VLBI surveys which target previously known sources. Furthermore, the GOODS-North field is the only extragalactic legacy field with wide and deep ($\approx 10\,\mathrm{\umu Jy\,beam^{-1}}$) milliarcsecond resolution coverage by two VLBI networks (i.e., EVN and VLBA). This will enable three key tests, in the context of continued development of wide-field VLBI techniques. Firstly, it offers independent measurements enabling a statistical comparison of the astrometry and absolute flux calibration of the two instruments, therefore constraining the systematic uncertainties. Secondly, it enables a search for highly time-variable sources at milli-arcsecond scales. Finally, it serves an important technical role, in validating the wide-field VLBI survey design and calibration and imaging pipeline, as described in Paper I. 

As presented in Paper I, several new candidate sources were identified in the VLBA survey which are either not in the EVN survey footprint or are below its non-uniform sensitivity limits and require further characterisation.  These aforementioned criteria have a fundamental role in these series of papers based on this survey, particularly as a physically well-understood comparative sample for the statistical calibration, source probabilistic and source characterisation to be presented in a future publication and its possible extension. Furthermore, this work plays an informative role for decisions on wide-field survey strategies in the next-generation VLBI networks, including SKA-VLBI and the African VLBI Network. 

This paper is organized as follows. Section~\ref{sec:radioGN} describes the existing deep radio surveys of the GOODS-North field. Section~\ref{sec:results} compares the imaging data products and derived properties of several VLBI-selected sample from the VLBA, EVN, \emerlin{} and VLA. In section~\ref{sec:discussion} we provide a discussion on theses results, including the origin of the compact and extended radio emission. Finally, section \ref{sec:Summary} presents the conclusions. For this work, we adopt the $\Lambda$CDM cosmology parameters with $H\mathrm{_0} = 67.4$\,kms$^{-1}$\, Mpc$^{-1}$ and   $ \Omega_{\rm m} = 0.315$ \citep{Planck2020a}. The spectral index, $\alpha$ uses the convection $S_\nu \propto \nu^{\alpha}$, where $S_\nu$ is the integrated flux density.

\section{GOODS-North Field: The VLBA and Other Deep Radio Surveys}\label{sec:radioGN}
The GOODS-North field \citep{Dickinson2003, Giavalisco2004} covers $\sim\,160\,$arcmin$^{2}$ and is centered on the Hubble Deep Field-North \citep[HDF-N;][]{Williams1996}, coordinates 12$^\mathrm{h}$36$^\mathrm{m}$44\fs27 and Dec. $+62^{\circ}14\arcmin24\farcs48$. The GOODS-North has deep coverage by \textit{Spitzer} in the near infrared (NIR), \textit{Herschel} in the far infrared (FIR), \textit{Hubble} Space Telescope (HST) in optical, \textit{Chandra} in X-ray and VLA, \textit{e}-MERLIN, and EVN in radio. 

This paper is the second presenting new VLBA data of the GOODS-North field (Paper I introduces the survey; see Section~\ref{sec:vlba}). The distinct advantage of VLBI observations, such as those from VLBA, is their insensitivity to diffuse emission on larger scales, and hence provide a clean tracer of high brightness temperatures ($>10^{5}$\,K) radio emission, which primarily originates from AGN cores at high redshifts \citep[e.g.][]{Kewley2000,Middelberg2013}. Conversely, due to this insensitivity to low surface brightness emission, VLBI observations tend to completely resolve out any extended emission, such as star formation and AGN jet structures. Therefore, to fully characterize the high redshift radio source in extragalactic deep fields, both intermediate ($\sim$ arcsec) and high ($\sim$ mas) angular resolutions scales are required. The multi-scale suite of data presented in this work covers angular resolution from $\sim 1\,$arcsec to $\sim\,5\,$mas.  We analyse the VLBA-selected sources from GOODS-North field at $\sim$1.6\,GHz using multiple instruments, the VLA, \emerlin{} and EVN with arcsec--mas coverage. Further details of each of the archival surveys used in this study are given in Sections \ref{sec:emerlinVLA} and \ref{sec:EVN+VLBA}. An overview of the final sample is provided in Section~\ref{sec:final_sample}. 

\subsection{VLBA 1.6\,GHz Survey}\label{sec:vlba}
Paper I provides full details of the VLBA survey strategy, observations, reduction and source extraction techniques. Here we provide a brief summary. The VLBA GOODS-North field was observed on September 13--November 2 2013 across 12 epochs, with each epoch $\sim 2\,$hrs long. The 1.6\,GHz survey was centred on the HDF-N field at R.A. $12^\mathrm{h}36^\mathrm{m}55^\mathrm{s}$ and Dec. $+62^{\circ}14\arcmin15\arcsec$, covering 160 arcmin$^2$ in area. This VLBA survey is based on a quasi-uniform sensitivity coverage of the entire CANDELS \citep[][]{Koekemoer_2011, Grogin2011} region of the GOODS-North field. The multiple phase centre correlation strategy was employed covering 205 phase centres uniformly separated by a distance of $\sim 35\arcsec$. A 64\,k $\times$ 64\,k pixel image corresponding to an area of $1\farcm0 \times 1\farcm0$ with a primary-beam-corrected sensitivity of $\sim 10\--15\,\mathrm{\umu Jy\,beam^{-1}}$ was generated for each phase centre, followed by source extraction. A source was classified as a VLBI detection if; a) the VLBI signal-to-noise ratio, $\mathrm{SNR_{VLBI}}\geq 7$ or, b) $\mathrm{SNR_{VLBI}} \geq 5.5$ if the VLBI position was within $0\farcs5$ of a catalogued multi-wavelength source. A total of 24 sources were detected, all of which have a detection at previous $\sim$1.5\,GHz observations of VLA and/or \emerlin{} (surveys described below). 

\subsection{VLA and \emerlin{}: arcsecond to sub-arcsecond scale radio observations: }\label{sec:emerlinVLA}
The GOODS-North field has been extensively studied at L-band ($1\mbox{--}2\,\mathrm{GHz}$) using the VLA \citep{Richards2000, Morrison2010,Owen2018}. \citet{Richards2000} catalogued a total of 371 sources within 20$^{\prime}$ of the HDF-N with an average angular size of $1\farcs8$. \citet{Morrison2010} improved this survey by increasing the total integration time to 165\,hr and catalogued 1230 radio sources at 5$\sigma$ within the central 40$^{\prime} \times 40^{\prime}$ with a central root-mean-square (RMS) of $\sim 3.9\,\mathrm{\umu Jy\,beam^{-1}}$. The median angular size of the sources was calculated as $\sim 1\farcs2$.  \citet{Owen2018} re-observed the GOODS-North field at $\theta_\mathrm{res} \sim 1\farcs6$ with the upgraded VLA in the A-configuration for a total of 39 hr with $\sim$ 33\,hr on target. They catalogued 795 sources at 5$\sigma$ within the central $9^{\prime} \times 9^{\prime}$ region with a central RMS noise of $2.2\,\mathrm{\umu Jy\,beam^{-1}}$. For the purposes of the VLA imaging used in this work, we used the primary beam corrected VLA A-array wide-field image from \citet{Owen2018} cropped to the \emerge{} DR1 region and re-sampled to 15\,k $\times$ 15\,k pixels with a sensitivity of $2.0\,\umu$Jy$\,$beam$^{-1}$ \citep{Muxlow2020}.

At higher resolutions, the \textit{e}-MERLIN Galaxy Evolution (\textit{e}-MERGE) survey \citep{Muxlow2005,Muxlow2020} used \emerlin{} to produce a naturally weighted primary beam corrected $15^{\prime} \times 15^{\prime}$ image with a resolution of $\mathrm{\theta}_\mathrm{res} \sim$ 0${\mathrm\farcs}2$ and a central sensitivity of $2.81\,\mathrm{\umu Jy\,beam^{-1}}$. A total of 848 sources were catalogued at 4.5$\sigma$. For this work, we used the DR1 image and the source catalogue as described in DR1 \citep[][Thomson et al. in prep.]{Muxlow2020}. The \emerge{} survey also includes observations at C-band  \citep{Guidetti2013,Guidetti2017}. These C-band data were used to derive our spectral indices, presented in Table\ref{Tab: Tb}.   

\subsection{EVN data: previous milliarcsecond scale radio observations}\label{sec:EVN+VLBA}

The European VLBI Network has observed the GOODS-North field multiple times across the past 25 years. The first survey concentrated on the smaller Hubble Deep Field-North (HDF-N) region at 1.6\,GHz and detected three VLBI candidates within the central $3\farcm5$ region \citep{Garrett2001}. This survey demonstrated the power of VLBI in distinguishing between AGN and starbursts in faint highly obscured radio source populations, and paved way for wide-field VLBI imaging and characterization of the faint radio sky at sub-mJy regimes. This survey was followed up with the 1.4\,GHz global VLBI observations of the HDF-N and Hubble Flanking Fields (HFF) for a total of 36\,h with $\sim$ 19\,hr on-source integration time \citep{Chi2013}. This survey covered a total area of 201\,arcmin$^2$ (HDF-N and HFF) with a central RMS of 7.3\,$\umu$Jy\,beam$^{-1}$ at $\theta_\mathrm{res}\sim\,4\,$mas. A total of 12 compact radio sources were detected at 5$\sigma$. This added nine new sources to the previous \citet{Garrett2001} survey.
 
\citet{Radcliffe2018} built upon these two VLBI surveys by imaging across the entire primary beam of the EVN. The GOODS-North field was observed for 24 hours using the EVN at 1.6 GHz on 5--6th June 2014. This survey targeted a region of 15$^{\prime}$ in diameter centred on the HDF-N. This was supplemented by targeting bright VLA sources up to 20$^{\prime}$ from the pointing centre. The observing strategy employed in this survey used five pointing centres to widen the area covered. As the EVN is a heterogeneous array, the primary beam response is different for each antenna, with the largest and most sensitive contributing the most to the inner central (7\farcm5) region and so the sensitivity decreases quickly as you move from the pointing centre. Consequently, this survey has a non-uniform sensitivity coverage over the full FoV. The data were naturally weighted for optimal sensitivity with a central RMS of $\sim$9$\mu$Jy beam$^{-1}$. A total of 31 sources were catalogued at 6$\sigma$, 19 more sources than previously detected in the \citet{Chi2013} survey.

\subsection{Final VLBI Sample}\label{sec:final_sample}
\begin{figure*}
\centering
\includegraphics[height=22cm, keepaspectratio=true]{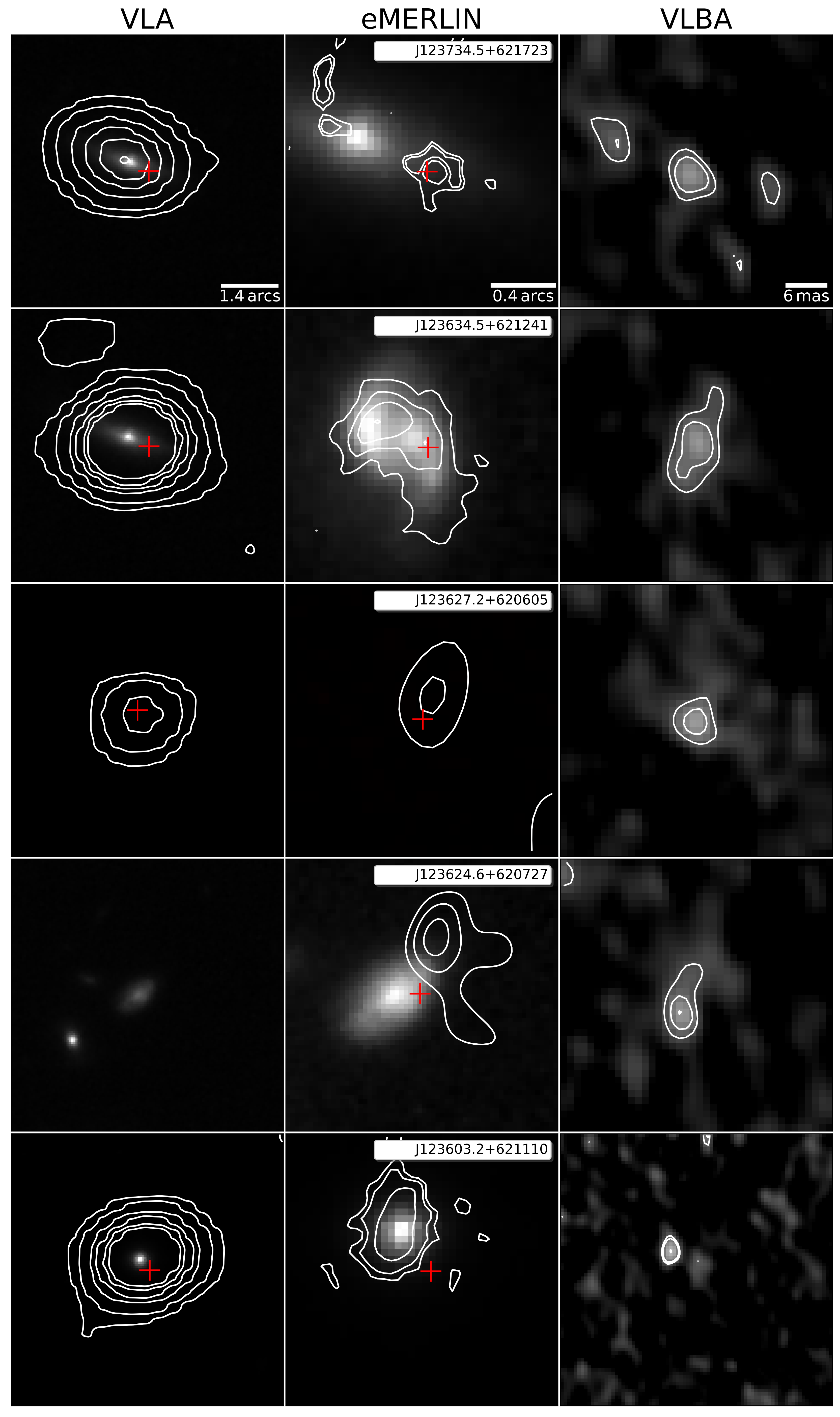}
\caption{Thumbnails of the five low-SNR ($\sim$5.5$\sigma$) new VLBI sources detected in the VLBA survey. Columns 1 and 2 show the contour maps for the sources in the VLA and \textit{e}-MERLIN respectively, at contour levels $3\sigma \times [-1,1,2,3,4,5,6,....,11]$ overlaying their 3D-\textit{HST} CANDELS optical counterparts in the F160W band, in grayscale. The VLBA contours are drawn at a central rms $\sim15\,\microJybm$ multiplied by factors of [2,3,5,7,9,11,13]. The red cross on the VLA and \emerlin{} images indicate the VLBI source position. The source J123627.2+620605 (3rd row) lacks both the optical counterpart and a redshift value. The source ID are shown on the \emerlin{} panel with the exact coordinates listed in Table\ref{table:Table2}.}
\label{figure:faintsources}
\end{figure*}

The 24 VLBA sources were cross-matched against the deep VLA and the \textit{e}-MERLIN source catalogues as described in Section~\ref{sec:emerlinVLA}. A total of 23/24 sources were found to be within a radius of $1\farcs0$ of a source catalogued in the VLA data and 23/24 were also matched to the \textit{e}-MERLIN. There was one non-matched source in each case. One source, J123701.1+622109 was only reported in the VLA catalogue \citep{Owen2018} and missing in the \emerlin{} catalogue, whilst the source J123624.6+620728 is only reported in the \textit{e}-MERLIN catalogue \citep{Muxlow2020} and not the VLA catalogue. We further compared the VLBA source catalogue to the EVN source catalogue \citep{Radcliffe2018} as described in Section \ref{sec:EVN+VLBA}. 18/24 sources were mutually detected across the 2 VLBI surveys. We note that, the 5/6 sources not detected in the EVN survey have the lowest signal-to-noise ratio in the VLBA survey, and were classified as VLBI detections using the $\mathrm{SNR_{VLBI}} \geq 5.5$ if the VLBI position was within $0\farcs5$ of a catalogued multi-wavelength source criteria (see Section~\ref{sec:vlba} and Paper 1). We refer to these 5 new VLBI detections as the low-SNR sources (or $\sim5.5\sigma$ sources; see Figure~\ref{figure:faintsources} and highlighted in Table~\ref{table:Table2}). The remaining source, J123746.6+621738 was in a region not covered by the EVN observations. Redshifts (either photometric or spectroscopic), were obtained for 23/24 targets from archival studies ranging from $0.321\--4.428$ with a median of $z = 1.145$. The source J123627.2+620605 has no available redshift. Table~\ref{table:Table2} presents a summary of the properties of the 24 sources, including their redshifts and their corresponding references, and the flux measurements across the different radio surveys.

In this work, we used this new VLBA sample in the GOODS-North field, to investigate the similarities and differences in the observed radio properties, between the two VLBI instruments (VLBA and EVN), and with a further comparison to the lower resolution data from the VLA and \textit{e}-MERLIN. This comparison is based on the radio properties of astrometric offsets, flux densities, source size distribution and brightness temperatures (presented in Section~\ref{sec:results}). These measurements are combined with other multi-wavelength data to characterise the AGN, and potential extended star-formation, in these targets in Section~\ref{sec:discussion}.

\section{Analysis and Results}\label{sec:results}

The positions, flux densities and angular sizes, deconvolved for the beam, of our VLBI sample of 24 sources, across the four instruments (see Section~\ref{sec:final_sample}) were measured by fitting a 2D Gaussian using the \textsc{casa} \textsc{imfit} task for the VLBA data and obtained from \citet{Radcliffe2018} for the EVN data, \citet{Owen2018} for VLA data and \emerge{} DR1 source catalogue from for the \emerlin{} data (\citealt{Muxlow2020}; Thomson et al. in prep) . These measurements enable a basic characterisation of the sources, and a comparison of the $\sim$1.6\,GHz radio emission across the different spatial scales as probed by the four instruments. In this section, we compare positions (Section~\ref{sec:positions}) and flux densities (Section~\ref{sec:fluxes}) between the VLBA measurements and counterparts of these sources in other surveys. We then present the results of the radio sizes and brightness temperatures as determined across the four instruments (Section~\ref{sec:tb}). These combined properties are used to discuss the origin of the radio emission, on different spatial scales, in Section~\ref{sec:discussion}. 

The images from VLBA, VLA and \emerlin{} are presented in Figure~\ref{figure:faintsources} and Figure~\ref{figure:montage}. The VLBA positions and flux density measurements across the four instruments, are presented in Table~\ref{table:Table2}. The restoring beam sizes, the major and minor axes of the deconvolved angular size (included the uncertainties as quoted by \textsc{casa imfit}), and their corresponding projected linear sizes are presented in Table \ref{Tab: Sourcesizes}. We note that some sources are completely unresolved, with no size information (labelled as ``Unresolved'' in the tables), whilst for some sources, a lower limit on the sizes was returned by the fitting routine. Table~\ref{Tab: Tb} presents the brightness temperature and luminosity measurements.

\begin{landscape}
\begin{table}
\caption{Properties of the VLBA sample sources.}
\centering
 
\begin{tabular}{lllllllllllll}
    \hline

    & & & & & VLBA & VLBA & VLBA & EVN & EVN & EVN & \emerlin{} & VLA \\
 Source ID & RA(J2000) & DEC(J2000) & Redshift& $z$ & S$_{\mathrm{peak}}$ & $S_\mathrm{int}$ & S/N & S$_{\mathrm{peak}}$ & S$_\mathrm{int}$  & S/N & S$_\mathrm{int}$ & S$_\mathrm{int}$  \\
 & & & $z$     &  type &($\microJybm$) &($\umu$Jy) & & ($\microJybm$)  & ($\umu$Jy)  & & ($\umu$Jy) & ($\umu$Jy) \\
          \\
    (1) & (2) & (3) & (4) & (5) & (6) & (7) & (8) & (9) & (10) & (11) & (12) & (13)     \\       
    \hline
 
\textbf{J123751.2+621919} & 12:37:51.2341 &	62:19:19.0132 & 1.1073$^{+0.11}_{-0.05}$ & P$^{a}$ & $120.0\pm{12}$ & $151.0\pm{25}$&  7.8 & $>$111.0 &>181 & 8.8 & $157.6\pm66.0$ & $157.9\pm5.4$  \\ 

J123746.6+621738 & 12:37:46.6708 &	62:17:38.5995 & 2.0997 & P$^{a}$  & $497.0\pm12$& $713.0\pm26$& 34.4 & $\mathrm{No\,\,data}$  & $\mathrm{No\,\,data}$ & $\mathrm{No\,\,data}$ & $1124.4\pm58.9$ & $1191.9\pm4.7$ \\

\textit{J123734.5+621723$^\ast$}& 12:37:34.4377 &	62:17:22.9336 & 0.6402 & S$^{b}$ & 74.3$\pm5.2$& 86.0$\pm38$& 5.6& $\mathrm{Undetected}$  & $\mathrm{Undetected}$  & $\mathrm{Undetected}$  & $95.7\pm51.0$ & $98.8\pm6.7$ \\

J123721.2+621129& 12:37:21.2535 &	62:11:29.9654 &$2.02 \pm0.06$ & P$^{a}$   & 209.5$\pm9.8$& 256.2$\pm20$& 17.7& 328.0$\pm38.3$& 364.0$\pm41.6$&  20.2 & $382.4\pm12.8$ & $367.8\pm12.1$ \\ 

J123716.7+621733&  12:37:16.6814 &	62:17:33.3151  & 1.146 & S$^{c}$ &  92.0$\pm10$& 166.0$\pm27$& 8.1& 150.0$\pm23.8$ & 269.0$\pm32.7$&  8.2 & $332.4\pm57.1$ & $337.9\pm11.1$  \\

J123716.4+621512&  12:37:16.3749	& 62:15:12.3451  & 0.559 & S$^{c}$ & 79.3$\pm4.3$& 107.0$\pm24$& 7.0& 125.0$\pm20.3$ &  125.0$\pm19.7$&  7.9 & $166.5\pm41.5$ & $175.4\pm5.9$ \\

J123713.8+621826&  12:37:13.8710 &	62:18:26.3003  & 4.428 & P$^{d}$  & 287.7$\pm8.5$& 441.0$\pm20$ & 21.6& 501.0$\pm56.8$ &  629.0$\pm69.4$&  25.6 & $617.0\pm52.5$ & $623.5\pm19.4$ \\

\textbf{J123701.1+622109}&  12:37:01.1040 &	62:21:09.6213 & 0.8 & S$^{c}$ & 179.2$\pm10$ & 218.3$\pm21$& 13.2 & $>$111.0  & $>$154.0&  11.5& $\mathrm{Undetected}$& $367.1\pm4.8$  \\

J123700.2+620909& 12:37:00.2475  &	62:09:09.7764 & 2.58$^{+0.07}_{-0.06}$ & P$^{a}$ & 103.0$\pm12$& 172.0$\pm31$& 8.7& 153.0$\pm23.4$ &  163.0$\pm24.1$&  9.4 & $296.2\pm10.3$ & $297.4\pm10.1$  \\

J123659.3+621832&  12:36:59.3345  &	62:18:32.5675  &2.167$^{+0.08}_{-0.07}$ & P$^{a}$  & 1912$\pm37$& 3270.0$\pm94$& 62.8&  2530.0$\pm328.9$ &  4430.0$\pm572.7$&  88.2 & $5034.1\pm22.1$ & $4600.8\pm78.0$ \\

J123652.9+621444&  12:36:52.8843  &	62:14:44.0703   & 0.321 & S$^{c}$  & $82.5\pm7.0$& 77.7$\pm12$& 7.8&  109.0$\pm15.1$  & 117.0$\pm15.6$ &  11.0 & $204.7\pm46.1$  & $214.0\pm7.8$ \\

J123646.3+621404&  12:36:46.3321  &	62:14:04.6929  & 0.961 & S$^{c}$ & 103.7$\pm7.6$& 144.0$\pm16$& 10.4&  191.0$\pm24.9$  & 192.0$\pm24.8$ &  12.3 & $273.2\pm67.8$  & $278.2\pm9.4$ \\

J123644.4+621133&  12:36:44.3876 &	62:11:33.1723  & 1.013 & S$^{c}$  & 243.6$\pm8.4$& 252.7$\pm14$& 21.9& 410.0$\pm44.8$   & 411.0$\pm44.7$ &  25.9 & $2228.3\pm44.9$ & $1792\pm76.2$   \\

J123642.1+621331&  12:36:42.0911 &	62:13:31.4305  & 2.018  & S$^{e}$ & 76.5$\pm7.5$ & 220.0$\pm40$& 8.9&  97.4$\pm18.0$   & 233.0$\pm27.9$ &  6.5 & $464.7\pm51.7$  & $461.3\pm6.9$ \\

J123640.5+621833&  12:36:40.5672 &	62:18:33.0800 &   1.146 & S$^{c}$ & 78.0$\pm5.2$& 190.0$\pm40$& 6.0& 141.0$\pm26.3$  & 141.0$\pm25.7$ &  7.5 & $314.1\pm48.5$ & $324.7\pm7.5$ \\

\textit{J123634.5+621241$^\ast$}& 12:36:34.4761  &	62:12:40.9562  &1.224 & S$^{c}$  & 57.6$\pm6.1$& 85.0$\pm14$& 5.7& $\mathrm{Undetected}$  & $\mathrm{Undetected}$  &  $\mathrm{Undetected}$  & $167.9\pm45.5$ & $188.5
\pm7.7$ \\

\textit{J123627.0+620605$^\ast$}& 12:36:27.2117 &	62:06:05.4395    &$\mathrm{No\,redshift}$  & --   &  85.2$\pm11$& 85$\pm23$& 5.6& $\mathrm{Undetected}$  & $\mathrm{Undetected}$  &  $\mathrm{Undetected}$  & $35.7\pm4.2$  & $34.3
\pm3.2$ \\

\textbf{\textit{J123624.6+620728$^\ast$}}& 12:36:24.5825 &	62:07:27.2875 & 0.5087 & S$^{b}$& 71.4$\pm11$& $86\pm27$ & 5.5 & $\mathrm{Undetected}$   & $\mathrm{Undetected}$   &  $\mathrm{Undetected}$ & $18.3\pm4.6$ & $\mathrm{Undetected}$ \\

J123623.5+621642&  12:36:23.5452	&  62:16:42.7450  &1.918 & S$^{f}$  & 166.4$\pm11$& 249.1$\pm25$& 12.2 & 222.0$\pm28.2$  & 383.0$\pm42.0$ &  12.8 &$402.0\pm61.9$   & $423.2\pm13.4$   \\

\textbf{J123622.5+620653}&  12:36:22.5101  &	62:06:53.8435  &1.94$^{\pm0.12}$ & P$^{a}$ & 164.1$\pm9.5$& 191.0$\pm19$& 10.7&  114.0$\pm19.0$ & 144.0$\pm21.4$& 8.2 & $257.8\pm10.3$  & $271.3\pm3.1$ \\

\textbf{J123620.2+620844}& 12:36:20.2633	& 62:08:44.2668   &1.0164 & S$^{b}$  & 76.9$\pm10$& 103.0$\pm22$& 6.0&   185.0$\pm25.8$& 185.0$\pm24.0$&  10.3 & $149.1\pm55.0$ & $126.4\pm4.1$ \\

J123617.5+621540 &  12:36:17.5563 &	62:15:40.7688  &1.993 & S$^{b}$  & 153.4$\pm10$& 209.1$\pm18$& 13.2&  177.0$\pm25.0$ & 192.0$\pm26.0$&  10.1 & $271.6\pm14.0$ & $240.3\pm8.7$ \\

J123608.1+621035& 12:36:08.1210 &	62:10:35.9089   &0.679 & S$^{c}$   & 99.3$\pm14$& 138.9$\pm37$& 7.8& 122.0$\pm16.8$ & 140.0$\pm18.2$&   11.1 & $218.6\pm40.3$ & $217.1\pm2.7$ \\

\textit{J123603.2+621110$^\ast$}& 12:36:03.2189 &	62:11:10.6180  & 0.638 & S$^{c}$ & 74.9$\pm10$& 55.7$\pm15$& 5.5& $\mathrm{Undetected}$  & $\mathrm{Undetected}$  &  $\mathrm{Undetected}$ & $129.8\pm39.0$ & $142.0\pm6.6$ \\
    \hline
    
\vspace{-0.2cm}\\    
\multicolumn{13}{l}{\footnotesize \textbf {Notes:} (1), (2) and (3) source ID and VLBA position in RA(J2000) and Dec(J2000). The sources italicised and highlighted with an asterisk are the low-SNR ($\sim5.5\sigma$) VLBA sources, which are not detected in EVN.}\\

\multicolumn{13}{l}{The sources highlighted in bold are 5/24 variable sources as reported in \citet{Radcliffe2019}. (4) redshift ($z$); redshift type, P for photometric redshifts and S for spectroscopic redshifts, while the letters denote}\\

\multicolumn{13}{l}{the references - $^a$\citet{Skelton2014}, $^b$\citet{Barger_2008}, $^c$\citet{Xue2016}, $^{d}$\citet{Yang2014}, $^{e}$\citet{Murpy2017}, $^{f}$\citet{Smail_2004}. (6), (7), (8), (9), (10) \& (11) show the peak brightness,}\\

\multicolumn{13}{l}{$\microJybm$; the integrated flux density, S$_{int}$ ($\umu$Jy); and signal-to-noise (S/N) across the VLBA and EVN. (12) \& (13) show the S$_{int}$ ($\umu$Jy) across \emerlin{} and VLA}\\


\end{tabular}

 \label{table:Table2}
\end{table}
\end{landscape}

\subsection{Positional offsets}\label{sec:positions}
\begin{figure}
\centering
\includegraphics[width=1\linewidth, keepaspectratio=true]{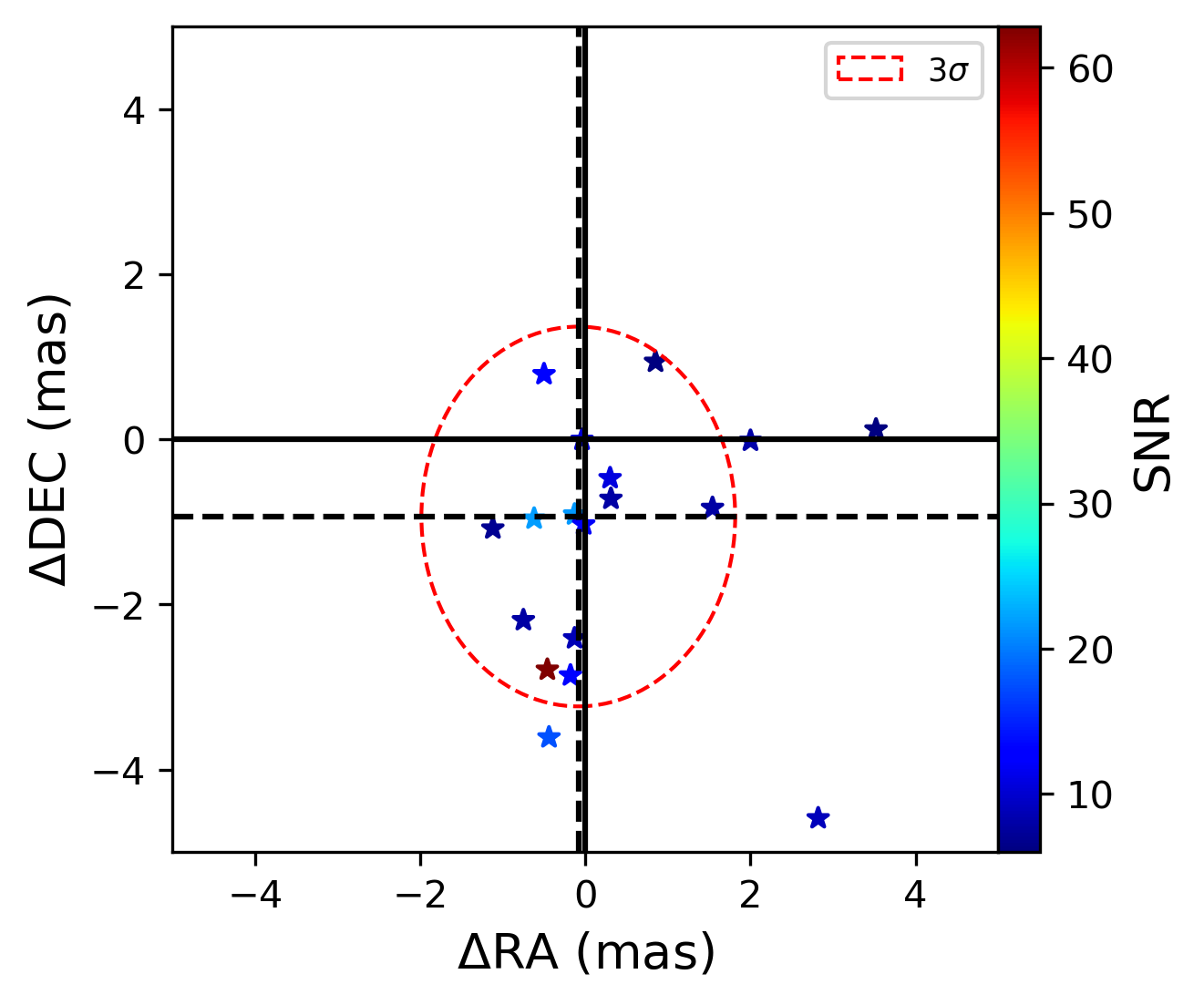}
\caption{The VLBI astrometric offsets, $\Delta$RA and $\Delta$Dec, between the VLBA and EVN for the 18/24 mutually detected sources. The solid black lines indicate $\Delta$RA = 0 and $\Delta$Dec = 0 mas while the dotted lines indicate the median astrometric offsets, indicating a good match between VLBA and EVN positions. The dashed red ellipse encloses sources within the offset scatter of 3$\sigma$ based on the individual $\Delta$RA and $\Delta$Dec values. The VLBA S/N is represented with the the colour scaling.}
\label{fig:VLBAEVNoffsets}
\end{figure}

The astrometric offsets between the VLBA and EVN data are shown in Figure~\ref{fig:VLBAEVNoffsets}, for the 18 sources with mutual detections. The median offsets between these positions are very small, with $\Delta$RA$\approx-0.085\,$mas and $\Delta$Dec $\approx-0.94$\,mas. The median astrometric offset between the two VLBI arrays is $<$1 mas, indicating an excellent agreement between the two data sets. The scatter in $\Delta$RA is $\approx\,1.3\,$mas and in $\Delta$Dec is $\approx\,1.5\,$mas. We note that the 4/18 sources showing the largest positional offsets, i.e., those beyond the defined 3$\sigma$ scatter (shown with the dashed ellipse in Figure~\ref{fig:VLBAEVNoffsets}), are some of the lowest S/N sources in the EVN.

\begin{figure}
\centering
\includegraphics[width=1\linewidth,  keepaspectratio=true]{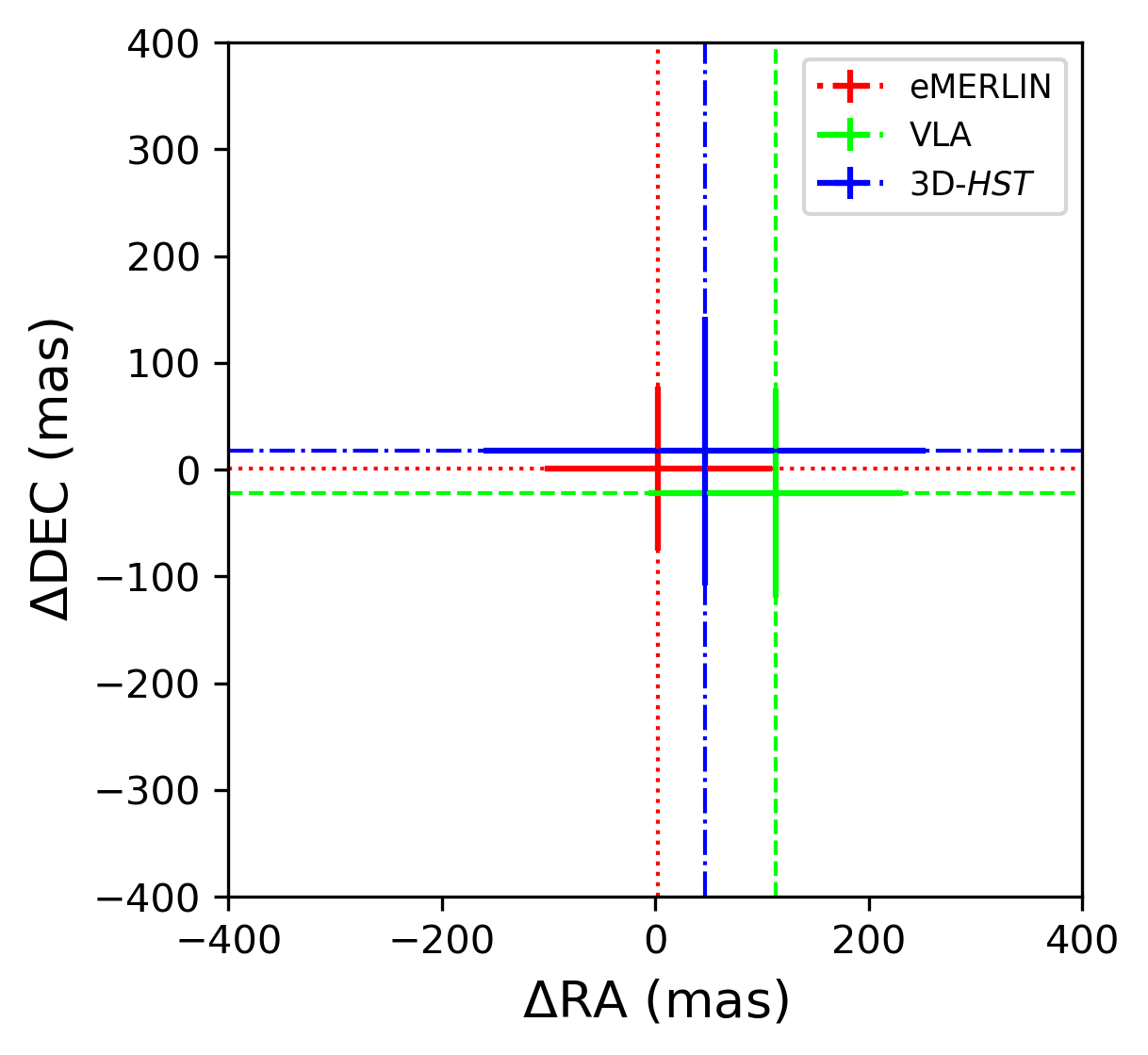}
\caption{The median astrometric offsets between the VLBA and \emerlin{} (red dotted line), VLBA and VLA (green dashed line), and VLBA and CANDELS 3D-\textit{HST} (dash-dotted blue line), plotted with respect to no offset to the VLBA position (0,0). The bold lines indicate the errors in the astrometric offsets, in the same respective colours.}
\label{fig:Othersastrometry}
\end{figure}

The median and scatter of the astrometric offsets between the VLBA positions and the \emerlin{}, VLA and optical CANDELS 3D-\textit{HST} \citep{Skelton2014} positions, are all represented in Figure~\ref{fig:Othersastrometry}. The median astrometric offsets between the VLBA and \textit{e-}MERLIN is $2.5\,$mas in $\Delta$RA and 1.2\,mas in $\Delta$Dec with a scatter of 106\,mas and 77\,mas, respectively. For VLBA and the VLA the median offsets are $\sim113\,$mas in $\Delta$RA and $-22$\,mas in $\Delta$Dec, with a scatter of 119\,mas and 98\,mas, respectively. Finally for, VLBA and CANDELS 3D-\textit{HST} \citep{Skelton2014} the median offsets are $\sim46\,$mas in $\Delta$RA and 18\,mas in $\Delta$Dec, with a scatter of 207\,mas and 126\,mas, respectively. 

All median offsets are consistent with (0,0) within the 1$\sigma$ scatter, which is to be expected, but also reassuring. A subset of sources do show large offsets between the VLBA and VLA and/or the VLBA and \text{HST} positions, which results in the relatively large scatter. However, such large offsets can arise in some sources due to different emission processes dominating the respective emission, and the relatively large differences in spatial resolution of the observations. The contribution of different processes (namely AGN and star-formation) to the radio emission on nuclear emission (probed by VLBA) compared to galactic scale emission (probed by VLA), is discussed in Section~\ref{sec:discussion}. 

\subsection{Flux density measurements}\label{sec:fluxes}
High angular resolution (mas) VLBI observations are insensitive to large scale and diffuse radio emission, including from extended AGN lobes, and diffuse star-formation-related emission. As such, we would expect that some fraction of the flux density observed at (sub-)arcsecond to remain undetected in the mas-scale VLBA images. Here we quantify this across our sample of VLBA-detected sources, by comparing the VLBA integrated flux densities with the VLA, \emerlin{} and EVN integrated flux densities. We derive the respective flux density ratios, which is parameterised as,
\begin{equation}
R_\mathrm{instrument}=S_{\mathrm{VLBA}}/S_{\mathrm{instrument}},
\end{equation} 
where instrument corresponds to the EVN, \emerlin{} or VLA. 

\begin{figure*}
\centering
\includegraphics[width=1\linewidth, keepaspectratio=true]{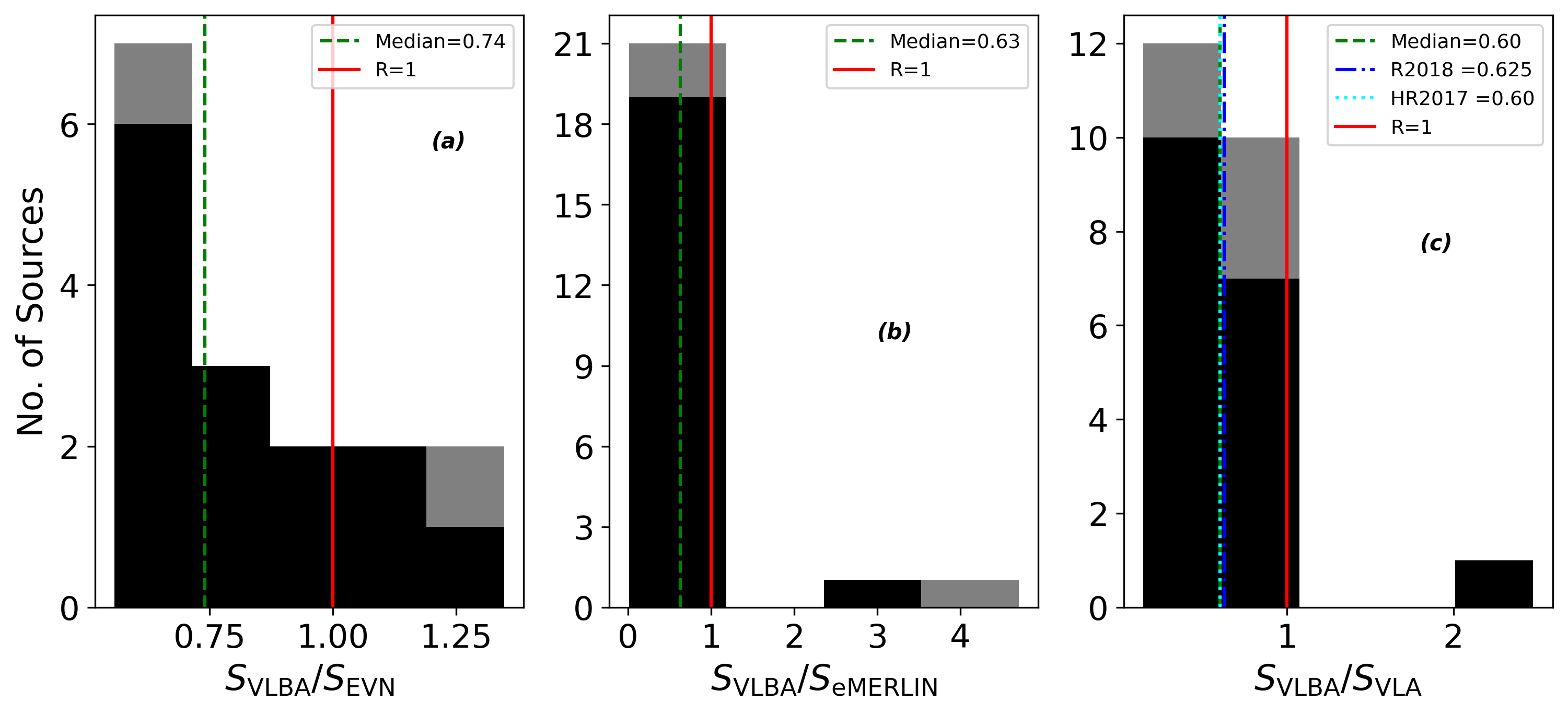}
\caption{Ratios of VLBA to the EVN, \emerlin{} and VLA flux densities. The green dashed lines indicate the median flux ratio $R_\mathrm{instrument}$, while the red solid lines indicate $R_\mathrm{instrument}$ =1. Panel (a) shows the $S_\mathrm{VLBA}/S_\mathrm{EVN}$ flux density ratios distribution between the VLBA and the EVN (for the 16 EVN sources with a flux density measurement), with a median of $R_\mathrm{EVN}= 0.74$. Panel (b) shows the $S_\mathrm{VLBA}/S_\mathrm{eMERLIN}$ flux density ratios with a median of $R_\mathrm{eMERLIN} = 0.63$. Panel (c) shows the source number distribution for the $S_\mathrm{VLBA}/S_\mathrm{VLA}$ flux density ratios with a median of $R_\mathrm{VLA} = 0.60$. The two blue lines in Panel (c), R2018 median = 0.625 and HR2017= 0.60 indicate the median flux density ratios of the GOODS-North VLBI source sample of \citet{Radcliffe2018} between the EVN and VLA, and the COSMOS VLBI source sample of \citet{HerreraRuiz2017} between the VLBA and VLA, respectively. The regions shaded in grey highlight the variable sources in this sample, as reported in the VLA variability study of the GOODS-North by \citet{Radcliffe2019}.}
\label{figure:Fluxratios}
\end{figure*}

Histograms of $R_\mathrm{instrument}$ flux density ratios are plotted in Figure~\ref{figure:Fluxratios}. We note that only sources with appropriate flux density measurements in the respective surveys are included in the histograms, that is 16/18 for EVN\footnote{2/18 EVN sources, J123751.2+621919 and J123701.1+622109, are excluded from this analysis since they do not have a reliable flux density measurement.} and 23 for each of VLA and \emerlin{} (see Table~\ref{table:Table2}). The sources which have previously been flagged as variable are highlighted in grey in the figure. 

For the comparison of EVN and VLBA flux density measurements (see Figure~\ref{figure:Fluxratios}, left panel a) we find median $R_\mathrm{EVN} =0.74$ with 12/16 sources showing $R_\mathrm{EVN}\,<\,1$. Only 2/4 of the remaining galaxies, J123640.5+621833 and J123622.5+620653, show significantly greater flux density in VLBA at $R_\mathrm{EVN} = 1.35$ and 1.33, respectively. The source J123622.5+620653 is also reported as a variable in \citet{Radcliffe2019}. For the remaining two sources, J123700.2+620909 $R_\mathrm{EVN}\,=\,1.06$ and J123617.5+621540 $R_\mathrm{EVN}\,=\,1.09$. We note all 16/18 sources agree between the total flux density measurements within $\sim$25\%. Since both the VLBA and EVN are probing similar spatial scales, one may also expect that $R_\mathrm{EVN}\,\approx\,1$, where there is no source flux density variability. However, there are differences in baselines configuration within the two arrays. The EVN has a larger fraction of relatively compact baselines, which also include large and most sensitive telescopes such as Effeslberg and Lovell, and is thus expected to be slightly more sensitive to diffuse and extended emission as compared to the VLBA, including some sources with a slighter larger beam. Additionally, we investigated the EVN/VLA flux ratios for the 13/31 (non-VLBA) sources reported in \citet{Radcliffe2018}. The median $R_\mathrm{VLA}\,\sim\,0.5$, lower than EVN/VLBA sources. These are also the brightest sources in the VLA sample, further evidence that the EVN is more sensitive to more resolved sources than the VLBA. While this is to be expected, its observational confirmation with different interferometric arrays is an important demonstration, particularly for the faint radio source population.

The comparison between VLBA and \emerlin{} (see Figure~\ref{figure:Fluxratios}, middle panel b), reveals a median flux ratio of $R_\mathrm{eMERLIN} = 0.63$, with two low-SNR ($\sim5.5\sigma$) sources (J123627.2+620605 and J123624.6+620728) showing $R_\mathrm{eMERLIN}>1$. The source J123624.6+620728 is reported as variable in \citet{Radcliffe2019} and remains undetected in the \citet{Owen2018} VLA source catalogue, while the source J123627.2+620605 shows a similar trend in VLA with $R_\mathrm{VLA}>1$. The median value $R_\mathrm{eMERLIN} = 0.63$ is consistent with expectations, due to the sensitivity to larger scales structures than the VLBA, we expect $R_\mathrm{eMERLIN}<1$. Whilst due to the longer baselines of \emerlin{} compared to VLA, one would anticipate $R_\mathrm{eMERLIN}>R_\mathrm{VLA}$, as is the case. 

The majority of the VLBA-VLA flux density ratios (see Figure~\ref{figure:Fluxratios}, right panel c) show $R_\mathrm{VLA}<1$, with only one source (J123627.2+620605) showing $R_\mathrm{VLA}>1$. This is a $\sim 5.5\sigma$ source and shows the largest $R_\mathrm{VLA}=2.48$. The median $R_\mathrm{VLA} =\,0.6$ value is consistent with other similar surveys. Both the VLBA-COSMOS survey \citep[]{HerreraRuiz2017} and the EVN-GOODS-North survey \citep[]{Radcliffe2018} find $R_\mathrm{VLA}\sim 0.6$. This indicates that, on average, $40$\% of the large-scale radio flux is resolved out in such VLBI-selected samples.

\subsection{Radio sizes and brightness temperature}\label{sec:tb}
At $\theta_{\rm res} \sim 1\farcs5$, the VLA observations correspond to a physical scale of $\sim 12\,$kpc at $z \sim 2$. On the other hand, VLBI mas angular scales, $\theta_{\rm res} \sim 5$\,mas, offer spatial resolutions $\sim 50$\,pc at $z\sim 2$. Therefore, the data presented here enables a comparison from arcsecond to mas imaging scales for extended emission and compact radio structures. The radio sizes of the sources across the four instruments (for which the required data are available) are presented in Table~\ref{Tab: Sourcesizes}.

\begin{figure}
\centering
\includegraphics[width=1\linewidth, keepaspectratio=true]{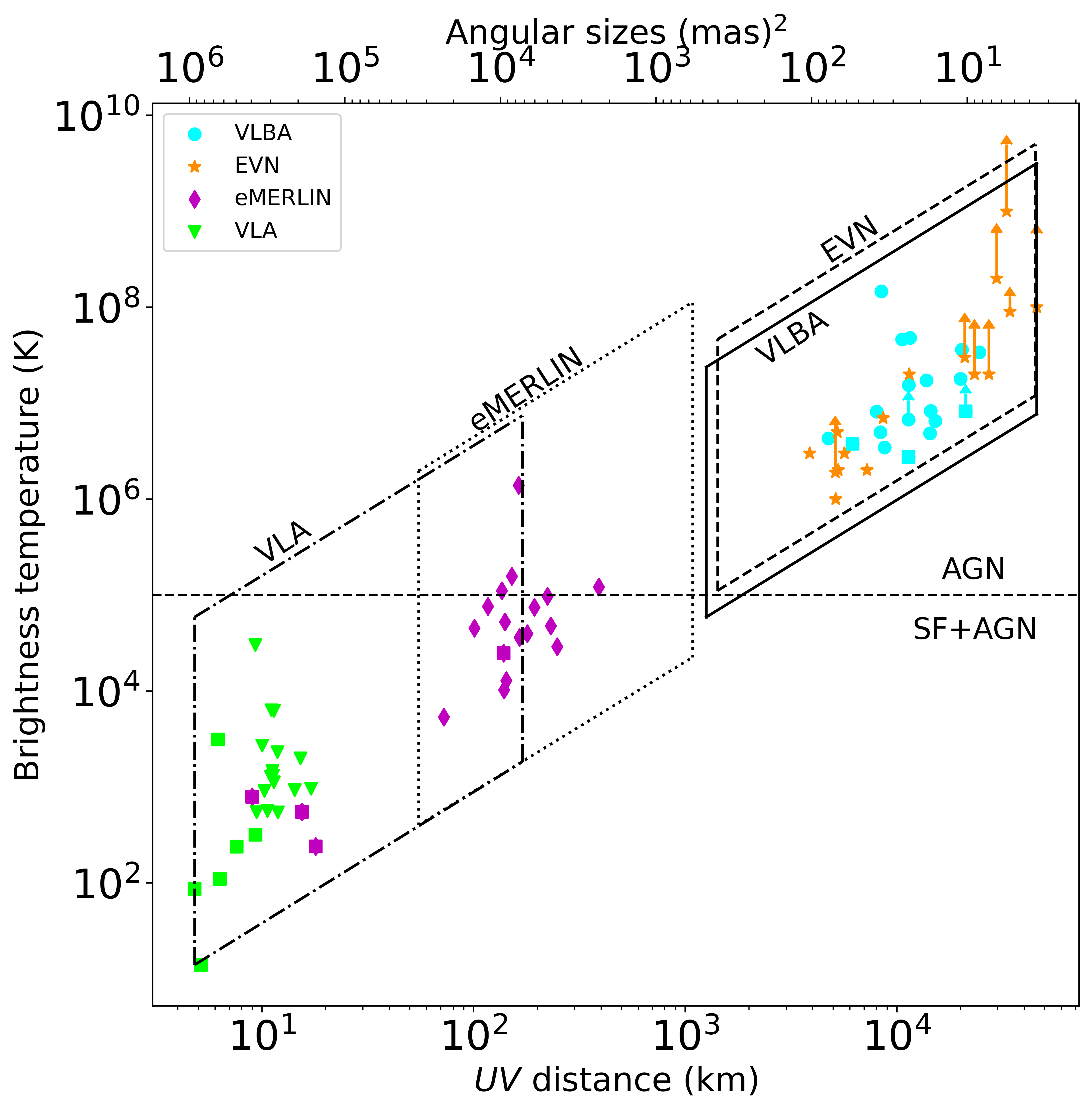}
\caption{The brightness temperature $T_\mathrm{B}$ (K) distribution across the VLBA, EVN, \textit{e}-MERLIN and the VLA as a function of both the $uv$-distance (km) and the angular sizes (mas)$^2$. The squares highlight the low-SNR ($\sim$5.5$\sigma$) sources. The error bars on the VLBA and EVN data are indicative of lower limits on the brightness temperature and upper limits on the deconvolved angular size measurements. The parallelograms highlight the minimum and maximum brightness temperatures limits that should be ideally recovered in each survey by applying equations \ref{TBmin} and \ref{TBmax} from \citet{Lobanov2015}. The black dotted line indicate the $\mathrm T_{B} > 10^{5}$\,K threshold typically used to separate AGN and star formation powered radio emission \citep{Condon1991}. Applying VLBI observations resolves bright compact (AGN) cores that would otherwise remain unresolved in both the \emerlin{} and VLA  and as such mis-classified as star formation.}
\label{figure:Tb}
\end{figure}

For the VLBA measurements, 5/24 sources are unresolved, and therefore only have an upper limit on the source extent. The angular sizes of the other sources range from 20--88\,parsec, 7--113\,parsec in the EVN, 0.2--4.4\,kpc in \textit{e}-MERLIN and  2--10\,kpc in VLA. As expected, these sources have typical physical sizes on 10s\,parsec scales in both the VLBA and the EVN, while the \textit{e}-MERLIN sizes lie typically at $\sim$1\,kpc scale and the VLA sizes on a few kiloparsec scales \citep[e.g.][]{Muxlow2005,Morrison2010,Barger2017,Radcliffe2018}. 

We can combine the measured flux densities and sources sizes, across the four different instruments to determine the measured brightness temperatures, $T_\mathrm{B}$. The brightness temperature can be used to determine whether the detected radio emission is likely to be dominated by nuclear AGN emission (highest brightness temperatures). We derive the brightness temperatures for our VLBI sample by assuming that the brightness temperature distribution of a source at a given redshift $z$ can be modelled as an elliptical Gaussian distribution as described in \citet[][]{Condon1982,Radcliffe2018}:
\begin{equation}\label{TBequation}
T_\mathrm{B} = 1.22 \times 10^{12}(1 + z) \left (\frac{S_\nu}{1\,\mathrm{Jy}}\right)\left(\frac{\nu}{1\,\mathrm{GHz}}\right)^{-2}\left(\frac{\theta_\mathrm{maj}\theta_\mathrm{min}}{1\,\mathrm{mas^2}}\right)^{-1}~\mathrm{K},
\end{equation}
where $\mathrm{\theta_{maj}}$ and $\mathrm{\theta_{\mathrm min}}$ are the deconvolved major and minor axes from the elliptical Gaussian model, $S_{\nu}$ is the observed flux density, and $\nu$ is the observing frequency. The derived $T_\mathrm{B}$  across the VLBA, EVN, \emerlin{} and VLA are shown in Table \ref{Tab: Tb}.

The brightness temperature range across the VLBA is $>10^6$--$10^8$\,K, EVN is $>10^6$--$10^9$\,K, \emerlin{} is $10^2$--$10^6\,$K and VLA is $10^1$--$10^4\,$K. Figure~\ref{figure:Tb} shows the brightness distribution of the VLBI sample as a function of the angular sizes across the four instruments.

The measured brightness temperatures cover orders of magnitude ranges, for the same sources, when measured across the different instruments. This is expected, based upon the \textit{uv}-coverage of the respective instruments. To show this explicitly, we calculate the probed brightness temperature ranges as a function of both \textit{uv}-coverage (per instrument spatial resolutions) and the recovered source sizes. We deduced the minimum and maximum brightness temperature for our interferometric visibilities for each array (assuming point source emission) and apply the minimum and maximum brightness temperature equations as discussed and derived by \citet{Lobanov2015}:

\begin{equation}\label{TBmin}
T_\mathrm{b,min} = 3.09 \left(\frac{B}{\mathrm{km}}\right)^{2} \left (\frac{V_{q}}{\mathrm{mJy}}\right) \mathrm K
\end{equation}
and
\begin{equation}\label{TBmax}
T_\mathrm{b,max} = 1.14 \left (\frac{V_{q}+\sigma_{q}}{\mathrm{mJy}}\right) \left(\frac{B}{\mathrm{km}}\right)^{2} \left (\mathrm{ln} \frac{V_{q}+\sigma_{q}}{V_{q}}\right)^{-1} \mathrm K.
\end{equation}

T$_\mathrm{b,min}$ is the minimum brightness temperature that can be observed given both the minimum and maximum baseline lengths, $B$ and $V_{q}$ is the visibility amplitude.T$_\mathrm{b,max}$ is the maximum measurable brightness temperature that should be detected at the minimum and maximum baselines of an array, while $\sigma_{q}$ is the sensitivity of the array. The derived minimum and maximum brightness temperatures regions covered by the VLA, \emerlin{}, EVN and VLBA are highlighted by the parallelograms on Figure~\ref{figure:Tb}.  

The high values of $T_\mathrm{B} > 10^{5}$\,K across both the VLBA and EVN is a strong indicator of the presence of AGN-dominated radio emission and confirms that our VLBI selection criteria returns a sample of AGN systems, based on brightness temperature \citep{Condon1991,Middelberg2008}. We also expect all of the unresolved sources to be AGN dominated (see parallelograms in Figure~\ref{figure:Tb}). On the other hand, the scales probed by the \emerlin{} observations are almost equally sensitive to nuclear AGN activity, and extended emission which could be a mixture of star formation and other AGN processes (e.g., jets/lobes and outflow-induced shocks). We discuss the contribution of point like and extended radio structures in Section~\ref{sec:discussion_extended}. Across our sample, only a small fraction could be confirmed to be AGN-core dominated based upon the \emerlin{} brightness temperature measurements. On the scales probed with the VLA, it is even more difficult to confirm an AGN origin of the emission based upon brightness temperatures alone. Indeed, none of our sample have $T_\mathrm{B} > 10^{5}$\,K. Of course, this does not rule out AGN origin (jets or shocks) of the radio emission measured in the VLA data, as we discuss in Section~\ref{sec:discussion}.

\section{Discussion}\label{sec:discussion}

In the previous section we presented a comparison of the positions, flux densities, and brightness temperatures of the 24 VLBI-selected sources from our new VLBA survey of the CANDELS GOODS-North field (see Paper I). This sample reaches low VLBI flux densities, including 5 new sources with low integrated flux densities from VLBA of $\sim$50--100$\mu$Jy, that were previously not recorded from the EVN observations of the same field (\citealt{Radcliffe2018}).  

VLBI observations as those carried out with our VLBA survey, are sensitive to brightness temperatures of $T_\mathrm{B}\sim 10^{6\mbox{--}10}\,$K (Figure~\ref{figure:Tb}), and these detections confirm our selected sample comprises of galaxies hosting a radio AGN core \citep[e.g.,][]{Condon1991, Middelberg2008,Radcliffe2018,Morabito2022}. In this section we discuss the nature of the AGN, including a comparison to other AGN tracers (Section~\ref{sec:discussion_core}) and also assess the nature of the extended radio emission in this sample (Section~\ref{sec:discussion_extended}). 

\subsection{Identifying and characterising the AGN}\label{sec:discussion_core}

Whilst radio emission benefits from being insensitive to obscuration, the origin of the radio emission can be attributed to either star-formation related processes, or AGN-related processes (including AGN cores, extended jets and shocked gas due to outflows; e.g., \citealt{Condon2012,Padovani2016,Panessa2019}). Therefore it is important to use multiple approaches to assess the origin of the identified emission on different spatial scales. The radio luminosities ($L_\mathrm{1.4GHz}$) of our sample, measured with four different arrays (see Table~\ref{Tab: Tb}), range between $L_\mathrm{1.4GHz} = 10^{23}$--$10^{27}$\,W\,Hz$^{-1}$ with a median of $\sim 10^{25}$\,W\,Hz$^{-1}$. A threshold of $L_{1.4\mathrm{GHz}}> 10^{24}$\,W\,Hz$^{-1}$ is sometimes used a boundary to classify sources as AGN 
(e.g., \citealt{Barger2017}). However, this relatively crude approach does not consider in detail the relative contribution to the radio luminosity from star formation versus AGN, and that these processes may be occurring on different spatial scales. 

It is possible to identify sources where the radio emission is in excess of that expected from star formation, due to the observed correlation between infrared emission and radio emission in star-forming galaxies \citep[e.g.,][]{Helou1985,Condon1992,Donley2005,Ivison2010,DelMoro2013}. Excess radio emission would indicate a dominant contribution from AGN. \cite{Radcliffe2021} provide a detailed characterisation for this method of EVN-detected VLBI AGN in the GOODS-North field. Therefore, here we only briefly explore this with our VLBA sample, which includes 6 new VLBI sources (see Section~\ref{sec:final_sample}), not included in the EVN sample. 

\begin{figure}
\centering
\includegraphics[width=1\linewidth, keepaspectratio=true]{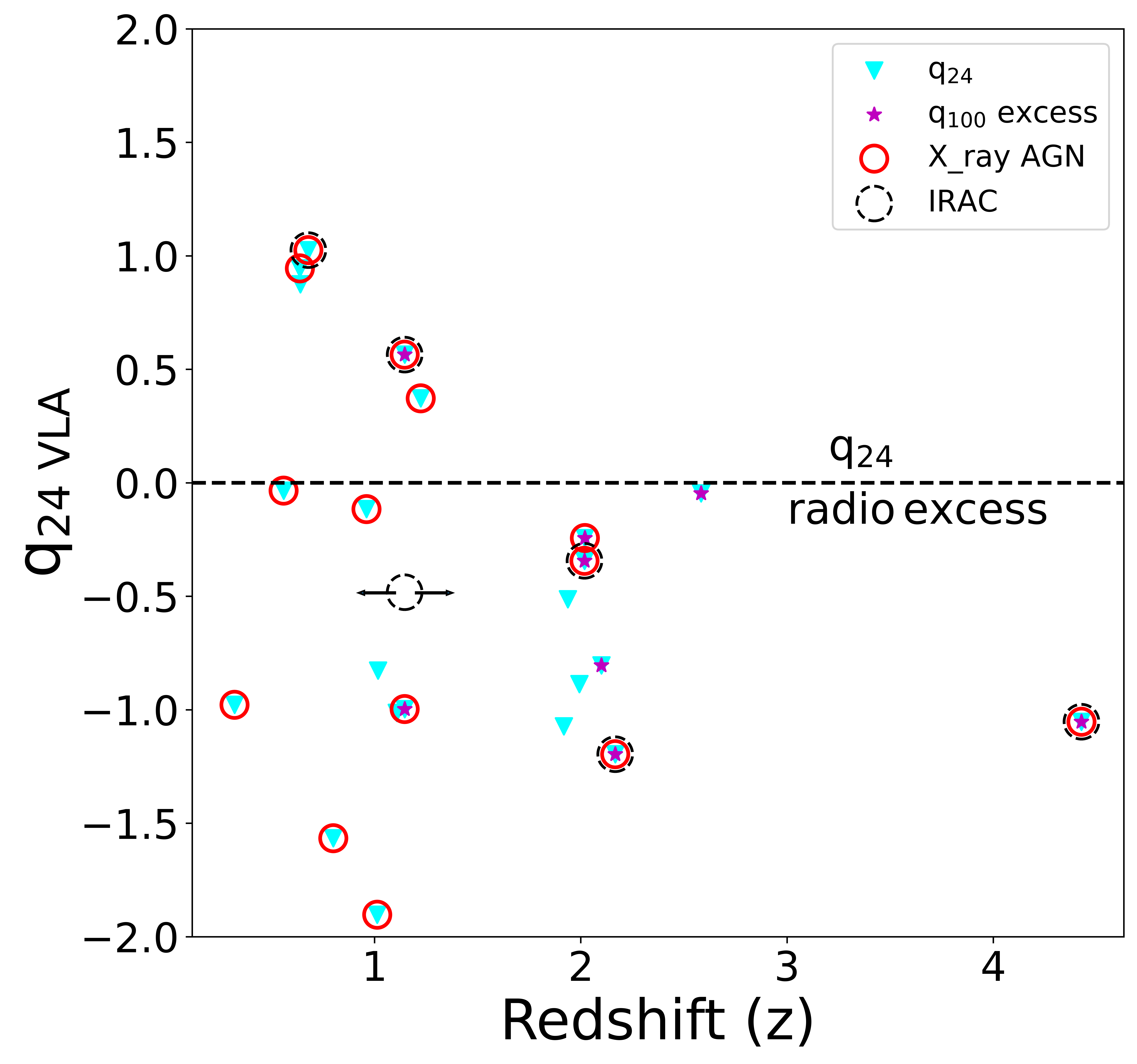}
\caption{The VLA-derived q$_{24}$ as a function of redshift for this VLBI-selected AGN sample. The q$_{24}<0$ cutoff for radio excess is represented with the dashed horizontal line. The red circles highlight the X-ray selected AGN, the dotted black circles highlight the IRAC selected AGN falling in the \citet{Donley2012} wedge while the purple star highlights the radio excess sources in q$_{100}$ for 8/12 sources with the required counterparts. The source shown with a black dotted circle with arrows lacks a redshift, and is also a low-SNR VLBA source.}
\label{figure:q24}
\end{figure}

We make use of the radio excess parameter, $q_{24} =\,\mathrm{log}(S_\mathrm{24\,\umu m}/S_\mathrm{20\,cm})$ and $q_{100} = \log(S_{100\,\mu m}/S_\mathrm{20\,cm})$ where $S_\mathrm{24\,\umu m}$ and $S_\mathrm{100\,\umu m}$ are the monochromatic infrared flux density at either $24\,\mathrm{\mu m}$ or $100\,\mathrm{\mu m}$ and $S_\mathrm{20\,cm}$ is the radio flux density at 1.4\,GHz \citep{Appleton2004}. The $S_\mathrm{24\,\umu m}$ data were obtained from \citet{Magnelli2011}, while the $100\,\mathrm{\mu m}$ data were obtained from \citet{Elbaz2011}. 23/24 targets are detected at $24\,\mathrm{\mu m}$ while only half (12/24) of the sample have appropriate detections at $100\,\mathrm{\mu m}$. In Figure~\ref{figure:q24}, we plot the $q_{24}$ values using the VLA flux density measurements against redshift. Using the \cite{Donley2005} radio excess selection criteria, where $q_{24}\,<\,0$ is classified as radio excess, 18/23 ($\sim 78$\,per\,cent) of our VLBI sample (with VLA detections and a redshift) show a radio excess. This fraction remains similar to the 79\,per\,cent found in \cite{Radcliffe2021}. Additionally, following the \cite{DelMoro2013} radio excess criteria which classify all galaxies below $q_{100}=1.5$ as showing radio excess, at least out to $z=2$, 8/12 (67\,per\,cent) of our VLBI sources with FIR counterparts show radio excess. We note that 3/5 of the sources not classified as radio excess are low-SNR ($\sim 5.5\sigma$) sources, highlighting that these weaker sources are less likely to be classified as AGN dominated using this approach 

Radio emission is able to identify AGN which are either radiatively efficient (sometimes referred to as `quasar mode' or `radiative AGN'), or radiatively inefficient (e.g., \citealt{Heckman14,Hardcastle2020}). In the GOODS-North sample of \citet{Guidetti2017}, $33\,$per\,cent of our sources are classified as radiatively efficient AGN while $\sim 61\,$per\,cent are classified as radiatively inefficient AGN. It is therefore interesting to assess the fraction of VLBI-detected AGN which are also classified as AGN based upon their X-rays or infrared colours, which are both tracers of radiatively efficient AGN. We find 14/24 (58\,per\,cent) sources are classified as X-ray AGN based on the catalogues of \citet{Yang2014} and \citet{Xue2016}. Assuming the IRAC "AGN wedge" selection of \cite{Donley2012}, we find only 6/24 (25\,per\,cent) sources to be selected as AGN. We highlight these X-ray and IRAC-selected AGN in Figure~\ref{figure:q24}. Again, these detection rates are consistent with the EVN VLBI sample of \cite{Radcliffe2021}, who perform a more complete analysis using multiple AGN selection methods. 

The 4/5 sources classified as non-radio excess $q_{24}>0$,(J123716.7+621733, J123634.5+621241, J123608.1+621035 and J123603.2+621110) are X-ray selected AGN in \citet{Xue2016}, with two sources (J123716.7+621733 and J123608.1+621035) further classified as AGN in the IRAC colour-colour diagrams and the source J123716.7+621733 showing radio excess at $q_{100}\sim1.1$. The remaining source J123734.5+621723, a low-SNR source, is not identified as AGN using any of our multiwavelength analysis.

We confirm that the milli-arcsecond scales provided by VLBI are crucial in identifying embedded AGN cores in large-scale radio emission (see Figure~\ref{figure:Tb}; also see e.g., \citealt{Garrett2001,HerreraRuiz2017}). No other single AGN-selection method that we have explored is complete at detecting all of the VLBI AGN (see also \citealt{Radcliffe2021b}).

\subsection{Extended radio emission}\label{sec:discussion_extended}

With our VLBA measurements we can asses the nature of the extended radio emission. The VLBA-VLA flux density ratios show that $\sim60\,$per cent (on average) of the total emission in these sources can be attributed to the AGN cores (see Figure~\ref{figure:Fluxratios}); also see e.g., \citealt{HerreraRuiz2017,Radcliffe2018}. 

Based upon morphology it is challenging to identify the nature of the extended radio emission for distant sources, which could be tracing AGN-related emission (i.e.,  jets/lobes or shocks), extended star formation processes, or a combination of both. While collimated jets and lobes are an indicator of the presence of radio AGN on the largest scales, $\sim$kiloparsec-scale radio structures can remain featureless in radio images at $z\gtrsim1$ \citep[e.g.,][and the references herein]{Padovani2016}. Indeed the images presented in Figure~\ref{figure:faintsources} and Figure~\ref{figure:montage} reveal that at arcsecond scales (VLA images), most of our sources are compact and lack clear morphological structures. At the sub-arcsecond scales traced by \emerlin{}, our sources are mostly compact, with some evidence of slightly resolved structures which may be indicating the emergence of sub-kpc scale jets from the base of these compact objects.

\begin{figure}
\centering
\includegraphics[width=1\linewidth, keepaspectratio=true]{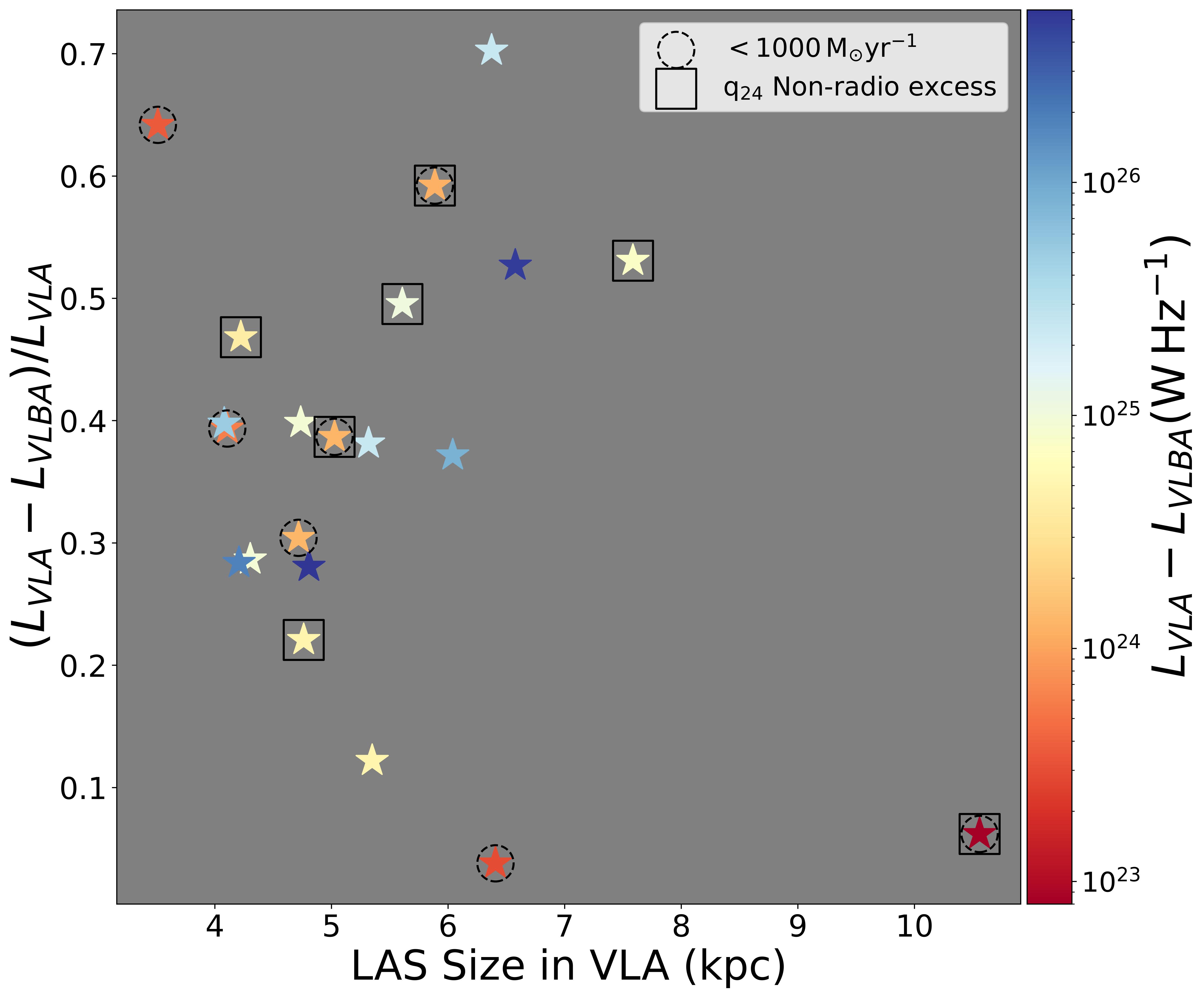}
\caption{The fraction of the extended radio emission luminosity as a function of the largest angular size in the VLA (for the 21 sources with VLA linear size measurements). The colours represent the luminosity of extended emission (i.e., with the VLBI core contribution subtracted). The dashed circles represent sources, where the inferred star formation rates (assuming all of the extended radio emission is due to star formation), would be $<$1000\,M$_{\odot}$\,yr$^{-1}$. The squares represent sources considered non-radio excess where $q_{24}>0$ in $\mathrm{log}(S_\mathrm{24\,\umu m}/S_\mathrm{VLA-VLBA})$. Only three non-radio excess sources show realistic SFRs$<$1000\,M$_{\odot}$\,yr$^{-1}$, for which the extended emission could be star-formation dominated based upon the $q_{24}$ parameter (highlighted in both squares and dashed circles).}
\label{figure:luminosity_LAS}
\end{figure}

We investigated the origin of the extended radio emission further by assessing the likelihood that it is associated with star formation. We explore the extended (non-core) radio emission only, by subtracting $L_\mathrm{VLBA}$ from $L_\mathrm{VLA}$. In Figure~\ref{figure:luminosity_LAS}, we plot the fraction of radio luminosity associated with extended emission, as a function of the radio size from VLA (for the 21 sources with the required measurements, see Table~\ref{Tab: Sourcesizes}). The data points are colour-coded by $L_\mathrm{VLA}$-$L_\mathrm{VLBA}$. We find that $\sim$5--70\,per\,cent of the total radio luminosity is attributed to these extended scales, but with a wide range of $L_\mathrm{VLA}$-$L_\mathrm{VLBA}$ values, ranging from $\sim$3\,$\times$10$^{23}$\,W\,Hz$^{-1}$ to $\sim$6\,$\times$10$^{26}$\,W\,Hz$^{-1}$. 

If we assume that all the extended (non-core) emission was attributed to star formation, we can calculate an inferred star-formation rate (SFR) using established relations for star-forming galaxies. Following \citet{Kennicutt2012} the inferred SFRs for half of our VLBI sample is $>5000$\,M$_{\sun}$\,yr$^{-1}$. Only 7 sources in our sample would imply potentially realistic SFRs of $<1000$\,M$_{\sun}$yr$^{-1}$ (highlighted as dashed circles in Figure~\ref{figure:luminosity_LAS}). SFRs of a few hundred M$_{\sun}$yr$^{-1}$ can be found in some of the most extreme star-forming galaxies at these redshifts \citep[e.g.][]{Santini2009,Popesso2023}. However, if we again employ the ``radio excess'' ($q_{24} =\,\mathrm{log}(S_\mathrm{24\,\umu m}/S_\mathrm{VLA-VLBA})$ approach to assess if this extended radio emission is beyond that expected from star formation, only 3/7 targets (J123734.5+621723, J123608.1+621035 and J123603.2+621110) considered non-radio excess where the extended radio emission is potentially dominated by star formation (see squares in Figure~\ref{figure:luminosity_LAS}). Of these three, two are from the low SNR part of the sample (i.e., they have lower brightness). This further indicates that by pushing to lower flux densities in VLBI, we are more likely to find hybrid objects in which the radio emission has both an AGN and a star-formation related component. The remaining 4/7 sources considered non-radio excess (J123716.7+621733, J123646.3+621404, J123634.5+621241 and J123622.5+620653), have high luminosities $L_\mathrm{VLBA}>10^{24.5}$\,W\,Hz$^{-1}$ and future work should be done to investigate these non-radio excess sources with high radio luminosity. In all of the other sources, the extended radio emission is likely to be dominated by AGN-related processes. 

As a final note on the low-SNR ($5.5\sigma$) VLBA sources (see Section~\ref{sec:EVN+VLBA}), whilst they remain undetected in the EVN GOODS-North survey (but were detected in both the \emerlin{} and the VLA) which employed a detection threshold of $6\sigma$, we have confirmed these sources as AGN (Figure~\ref{figure:Tb}). All the low-SNR VLBI detections have VLBA flux densities $< 100\,\umu$Jy and have mostly unresolved source morphologies on VLA scales (see Figure~\ref{figure:faintsources}). The source J123634.5+621241, is particularly interesting, in that there is extended \emerlin{} emission tracing the optical image, with potentially two radio peaks in the \emerlin{} data (see Figure~\ref{figure:faintsources}). This may indicate a merging system. However, only one of these peaks is detected in the VLBA image. This peak is clearly a radio AGN core. The classification of this source is one of the most debated issue in the GOODS-North field. While \citet{Kirkpatrick2012} and \citet{Muxlow2005} classify the source as a star-forming galaxy, \citet{Xue_2010} and \citet{Xue2016} classify the source as an X-ray AGN. \citet{Muxlow2020} classify the source as an extended merging starburst system undergoing intense star formation.

\section{Conclusions}\label{sec:Summary}
We have conducted a new VLBI 1.6\,GHz survey of the CANDELS GOODS-North field, using the VLBA. Paper I introduced the survey design, data processing, data products, and derived source counts. In this paper, we present new analyses of the 24 VLBA-detected sources, down to $5.5\sigma$ detection threshold, and which have radio counterparts in \emerlin{} and/or VLA. Of these, 6/24 are new VLBI sources in the GOODS-North field, without previous recorded detections from the EVN data of \citet{Chi2013} and \cite{Radcliffe2018}. Of these 6 new VLBI sources, 5/6 sources are low-SNR ($\sim 5.5\sigma$) detections, classified as VLBI-selected AGN based on available multiwavelength data. 

We studied the radio properties of these 24 sources across four different instruments (VLBA, EVN, \emerlin{} and VLA), covering a large range of probed spatial scales ($\theta_\mathrm{res}\approx0\farcs005\--1\farcs5$) with high sensitivity (rms$\approx2\--10\,\microJybm$). We further combined this radio analysis with multi-wavelength information to probe the nature of these VLBI-selected sources, on nuclear scales (i.e., a few tens of parsec) and galaxy scales (i.e., a few kiloparsec). Our primary conclusions are as follows:

\begin{itemize}
\item Comparing the VLBA and VLA flux density measurements, we find a medium ratio of $R \sim 0.6$ for $S_\mathrm{VLBA}/S_\mathrm{VLA}$ (Figure~\ref{figure:Fluxratios}). This is in  agreement with measurements from other similar VLBI surveys. This implies that, on average, $\sim60$ per cent of VLBI-selected sources (at these detection limits), can be attributed to compact VLBI cores. 
 \item The VLBA sources, have the expected high brightness temperatures ($T_{B}\gtrsim10^{6}$\,K), indicative of nuclear AGN emission. None of the 24 sources would have been reliably classified as an AGN based on the $T_{B}$ measurements from VLA, and only a small number ($\sim 30\,$per\,cent) would be candidates based on the \emerlin{} measurements (Figure~\ref{figure:Tb}). Of these VLBI-AGN, 15/24 are classified as (radiative) AGN based upon existing X-ray and infrared data. Combined, this highlights the importance of VLBI studies for characterising a complete AGN population, and the need to continue to expand the area coverage of sensitive VLBI surveys over extragalactic fields.
 \item To assess the origin of only the larger-scale radio (i.e., the non-AGN core) emission on a few sub-kpc ($\gtrsim100\,$pc) scales, we subtract the VLBA from the VLA luminosities for the 21 sources with linear size measurements from VLA. This emission contributes $\sim$5--70\,per\,cent of the total luminosity. Based upon the ``radio excess'' parameter and realistic star formation rates of $<1000$\,M$_{\sun}$\,yr$^{-1}$, only three sources could be star-formation dominated (Figure~\ref{figure:luminosity_LAS}). We conclude that, for the other 18 sources, the extended radio emission is likely to be AGN-dominated (i.e., tracing kpc-scale radio jet/lobes or shocks from outflows).

\end{itemize}
We have used VLBA and archival radio data to identify both AGN cores, and to assess the origin of the larger scale radio emission in sources at $z \sim$ 0.3--4. This work already provides SKA2-like spatial-scale coverage, which is useful for key technical verification for SKA-pathfinders such as astrometry and flux density scaling \citep[e.g.][]{Paragi2015}. Our current results may be observationally biased towards sources with detections of both compact emission and larger-scale radio emission. Specifically, VLBI-selected sources are skewed towards high brightness temperature sources and remain insensitive to low surface brightness sources. Some of the arising issues include determining the intermediate spatial scales covering the transition between pure AGN emissions and star-formation processed, could be addressed in future wide-field VLBI survey with an intermediate resolution ($\sim 10 - 100$\,mas) view of these faint sources.

\section*{Acknowledgements}
We thank the referee for their constructive comments that helped improve the quality of this manuscript. Research reported in this publication was supported by a Newton Fund project, DARA (Development in Africa with Radio Astronomy), and awarded by the UK’s Science and Technology Facilities Council (STFC) - grant reference ST/R001103/1. RPD acknowledge funding by the South African Research Chairs Initiative of the DSI/NRF. JFR acknowledge funding from the South African Radio Astronomy Observatory (SARAO), which is a facility of the National Research Foundation (NRF), an agency of the Department of Science and Innovation (DSI). AN and CMH acknowledges funding from an United Kingdom Research and Innovation grant (code: MR/V022830/1). The National Radio Astronomy Observatory is a facility of the National Science Foundation operated under cooperative agreement by Associated Universities, Inc. We acknowledge the use of the ilifu cloud computing facility – www.ilifu.ac.za, a partnership between the University of Cape Town, the University of the Western Cape, the University of Stellenbosch, Sol Plaatje University, the Cape Peninsula University of Technology and the South African Radio Astronomy Observatory. The Ilifu facility is supported by contributions from the Inter-University Institute for Data Intensive Astronomy (IDIA – a partnership between the University of Cape Town, the University of Pretoria, the University of the Western Cape and the South African Radio astronomy Observatory), the Computational Biology division at UCT and the Data Intensive Research Initiative of South Africa (DIRISA).

\section*{Data Availability}

Data underlying this article are publicly available in the NRAO \href{https://data.nrao.edu/portal/}{Archive Access Tool} under VLBA project code BD176. The reduced data will be shared on reasonable request to the corresponding author.


\bibliography{refs} 
\bibliographystyle{mnras}



\appendix{}

\section{Tables and thumbnail figures}
\begin{figure*}
\centering
\includegraphics[width=0.8\linewidth, height=24cm, keepaspectratio=true]{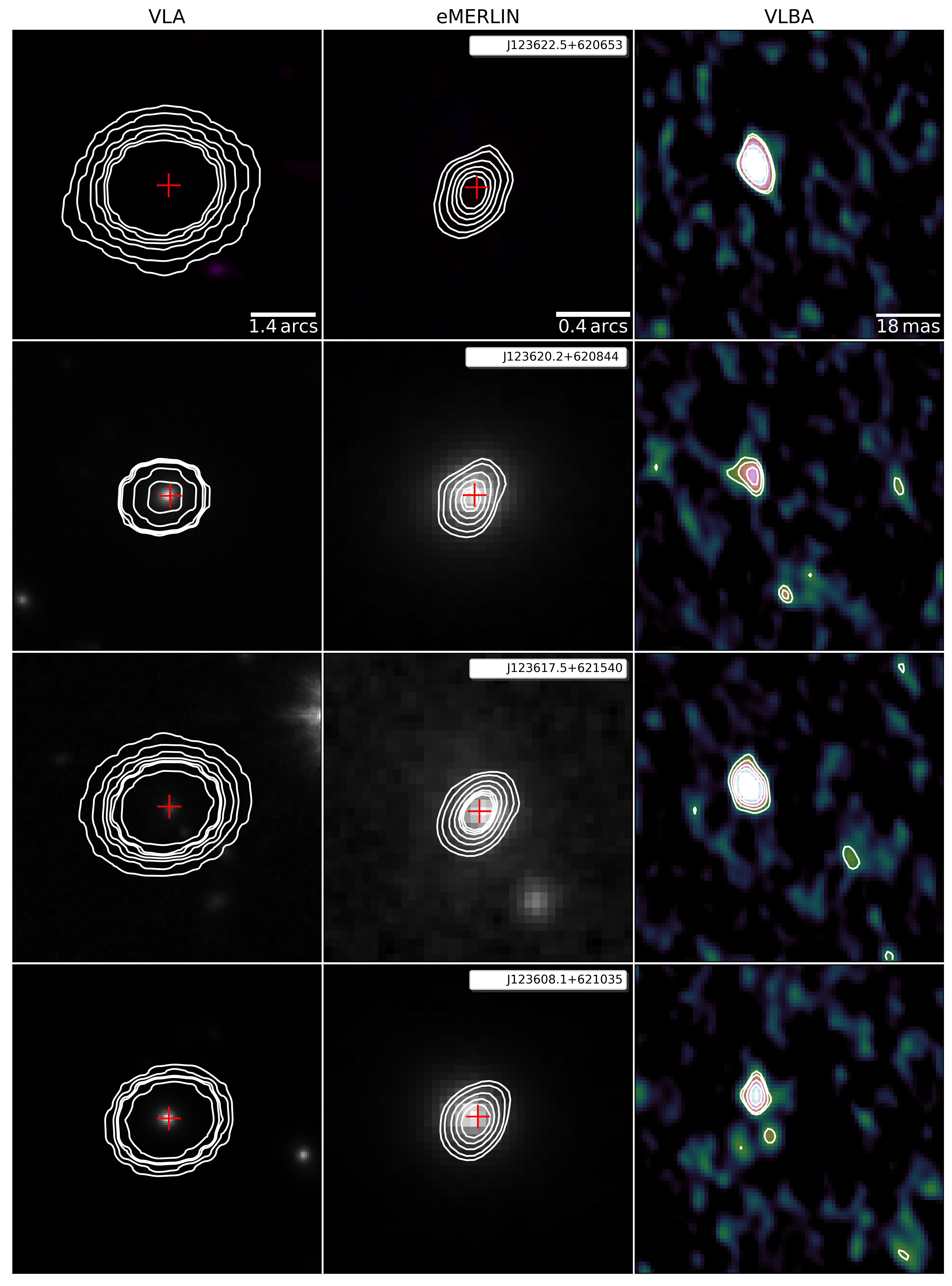}
\caption{A montage of the 19/24 VLBA CANDELS GOODS-North sources with $\mathrm{SNR_{VLBI}}\geq 7$. Columns 1 and 2 show the contour maps for the sources in the VLA and \textit{e}-MERLIN respectively, at contour levels $3\sigma \times [-1,1,2,3,4,5,6,....,11]$ overlaying their optical counterparts 3D-\textit{HST} CANDELS in the F160W band where available, in grayscale. The VLBA contours are drawn at a central rms $\sim15\,\microJybm$ multiplied by factors of [2,3,5,7,9,11,13]. The red cross on the VLA and \emerlin{} images indicate the VLBI position. The source ID are shown on the \emerlin{} panel with the exact coordinates listed in Table\ref{table:Table2}.}
\label{figure:montage}
\end{figure*}

Here we present all of the data tables for the 24 VLBA-selected sample in this work. Each table presents the relevant data for the measurements from VLBA, EVN, \emerlin{} and VLA. Table~\ref{table:Table2} provides an overview of the sample, including the flux density measurements. Table~\ref{Tab: Sourcesizes} presents the source sizes and corresponding restoring beams of each image. Table~\ref{Tab: Tb} presents the brightness temperature measurements. Figure~\ref{figure:montage} presents an overview of the 19/24 $\mathrm{SNR}_{VLBI}>7$ radio images. 

\begin{landscape}

\begin{table}
    \caption{Restoring beam sizes, angular source sizes (major axes $\times$ minor axes) in mas, and linear source sizes (major $\times$ minor axes) in parsecs across VLBA and EVN, and in kpc across the \emerlin{} and VLA. The restoring beam size for the \emerlin{} is $309\times213\,$mas while that of the VLA is $1.7\times1.4\,$arcsec. The $\sim 5.5\sigma$ (non-EVN detections) sources are italicized and highlighted with an asterik.}
    \label{tbl::agreement}
    \small
    \scalebox{1.2}{
    \centering
    \begin{tabular}{l ll llll llll}
        \hline
    \multicolumn{1}{l}{ }&\multicolumn{2}{c}{Beam Size}&\multicolumn{4}{c}{Angular size}& \multicolumn{4}{c}{Linear size}\\
    
        \multicolumn{1}{l}{ }&\multicolumn{2}{c}{(mas)}&\multicolumn{4}{c}{(mas)}& \multicolumn{2}{c}{(parsec)} & \multicolumn{2}{c}{(kpc)}\\
        \hline
        \textbf{Source ID}&\textbf{VLBA}&\textbf{EVN}&\textbf{VLBA}&\textbf{EVN}&\textbf{e-MERLIN}&\textbf{VLA}&\textbf{VLBA}&\textbf{EVN}&\textbf{e-MERLIN}&\textbf{VLA}\\
        \hline

J123751.2+621919&	9.1$\times$6.1&	11.9$\times$10.5 & <$8.2\times$2.9& Unresolved  &	175$\times$132&	762$\times$413&	<$68.9\times$24.4 & Unresolved & 1.5$\times$1.2&	6.4$\times$3.5\\

J123746.6+621738 &	9.0$\times$6.2& No\,data &  5.19$\times$4.49&  No\,data  &	 129.1$\times$75.9&	708$\times$418&		44.3$\times$38.3& No data &1.1$\times$0.7&	6.0$\times$3.6\\
 
\textit{J123734.5+621723$^\ast$} &	9.0$\times$6.0&	Undetected		& Unresolved & Undetected & 625$\times$246 & 1490$\times$620& Unresolved	 & Undetected & 4.4$\times$1.7	&10.6$\times$4.4\\

J123721.2+621129&	8.9$\times$6&	5.3$\times$4.5&			4$\times$2.7&	2.8$\times$<1.9&	110$\times$85&	502$\times$417&		34.3$\times$23.1&	24$\times$<16.5&	0.9$\times$0.7 &	4.3$\times$3.6\\

J123716.7+621733&	9$\times$5.9&	5.4$\times$4.6&			8.2$\times$4.4&	6.8$\times$5.1&	95$\times$56&	663$\times$436&		69.4$\times$37.2&	57.6$\times$43.2&	1.0$\times$0.5 &	5.6$\times$3.7\\

J123716.4+621512&	8.9$\times$5.9&	5.4$\times$4.6&			5.1$\times$3.4 &	10.4$\times$6.5&	111$\times$71&	617$\times$436&		33.9$\times$22.6&	69.1$\times$43.4&	0.9$\times$0.5 &	4.1$\times$2.9\\

J123713.8+621826&	9.1$\times$5.9&	5.3$\times$4.6&	 5.88$\times$4.45&	3.8$\times$<1.7&	95$\times$89&	618$\times$463&		40.0$\times$30.3&	28.5$\times$<12.9&	0.6$\times$0.6 &	4.2$\times$3.2 \\	

J123701.1+622109&	9.3$\times$5.8& $12.4\times11.0$  & $3.8\times2.9$ & 9.4$\times$7.3& Undetected& Unresolved &			29.3$\times$22.3& 72.9$\times$56.7& Undetected & Unresolved\\	

J123700.2+620909&	9.1$\times$5.9&	5.3$\times$4.5&			7.2$\times$5.3&	9.5	$\times$7.2&	85$\times$71&	497$\times$476&		59.1$\times$43.5&	78.3$\times$59.1 & 0.7$\times$0.6 &	4.0$\times$3.7\\

J123659.3+621832&	9.1$\times$5.8&	5.3$\times$4.5&			7.91$\times$4.51&	6.2$\times$<0.9&	121$\times$63&	566$\times$479&		67.2$\times$38.3&	52.3$\times$<7.7& 	1.0$\times$	0.5&	4.7$\times$4.5\\

J123652.9+621444&	8.8$\times$5.8&	14.8$\times$14.7& Unresolved	  &9.2$\times$4.8&	163$\times$58&	730$\times$532& Unresolved &44.1$\times$23.0&	0.8$\times$0.3&	2.5$\times$2.0\\

J123646.3+621404&	8.8$\times$5.8&	5.4$\times$4.5&	 		6.6$\times$2.6&	<$2.9\times$<2.5&	95$\times$50&	518$\times$421&		53.8$\times$21.2& 	<$23.9\times$<20.1&	0.8$\times$0.4&	6.4$\times$	3.4\\

J123644.4+621133&	9$\times$5.8& 5.3$\times$4.5&  Unresolved &	2.1$\times$<1.7&	165$\times$89&	772$\times$440& Unresolved &	17.6$\times $<13.9&	1.4$\times$	0.7&	6.4$\times$3.6\\

J123642.1+621331&	8.8$\times$5.9&	5.4$\times$4.5&	 		13.2$\times$5.9&	12.1$\times$8.5&	154$\times$78&	767$\times$443&		113.2$\times$50.6&	103.4$\times$73.2&	1.0$\times$0.5&	3.9$\times$3.5\\

J123640.5+621833&	9.1$\times$5.9&	5.3$\times$4.5&	18.8$\times$2.9& 12.3$\times$5.0&		103$\times$65&	560$\times$514&	$159.0\times24.5$&	104.4$\times$42.6&	0.9$\times$0.7&	3.9$\times$3.5\\

\textit{J123634.5+621241$^\ast$} &	8.9$\times$5.9& Undetected &  11$\times$<0.93	& Undetected & 894$\times$452&	888$\times$674&		94.0$\times$<7.9& Undetected & 7.6$\times$3.9&	5.9$\times$4.7\\

\textit{J123627.2+620605$^\ast$} &	9.4$\times$5.9& Undetected	& Unresolved	& Undetected &	Unresolved &	756$\times$603 &Unresolved & Undetected& Unresolved & $\mathrm{No\,Redshift}$ \\
 
\textit{J123624.6+620728$^\ast$} &	9.3$\times$5.9 & Undetected & 14$\times$1.7 & Undetected&	Unresolved& Undetected &  88.8$\times$10.8& Undetected & Unresolved &  Undetected \\	

J123623.5+621642&	8.9$\times$6.1&	5.4$\times$4.5&	 		5.4$\times$4.4&	5.9$\times$4.0&	106$\times$47&	617$\times$440&		46.6$\times$37.9&	50.9$\times$34.1&	0.9$\times$0.4&	5.3$\times$3.8\\

J123622.5+620653&	9.3	$\times$5.9	& $16.1\times15.3$&	 	4.4$\times$1.9& $13.2\times7.8$ &	103$\times$73&	553$\times$349&		37.9$\times$16.3& $113.7\times67.2$ &			0.9$\times$0.6&	4.8$\times$3.0\\	

J123620.2+620844&	9.2$\times$5.9&	5.3$\times$4.6&	 5$\times$3.2 &	<$3.2\times$<2.8& 	110$\times$<22&	 571$\times$338&		41.3$\times$26.4&	<26.3$\times$<22.9& 	0.9$\times$<0.2&		4.7$\times$2.4\\

J123617.5+621540&	8.9$\times$6.1&	5.4$\times$4.6&	 		5.9$\times$3.1&	3.7$\times$<2.8&	98$\times$40&	623$\times$478&		50.7$\times$26.6&	31.5$\times$<23.8&	0.9$\times$0.2&	5.3$\times$	4.1\\

J123608.1+621035&	9$\times$6.1&	16$\times$15.4&	 		10.6$\times$3.2&	11.1$\times$6.3&	<87$\times<$16&	692$\times$552&		77.0$\times$23.2&	80.8$\times$45.8&	0.7$\times$0.4 &	5.0$\times$3.9\\

\textit{J123603.2+621110$^\ast$} &	$9\times6.2$ & $\mathrm{Undetected}$ &	$\mathrm{Unresolved}$   & $\mathrm{Undetected}$ & $622\times$305 &	$832\times596$ & $\mathrm{Unresolved}$ & $\mathrm{Undetected}$ &		4.0$\times$2.4 & 	$5.9\times4.2$ \\

        \hline
    \end{tabular}
}
\label{Tab: Sourcesizes}    
\end{table}
\end{landscape}

\begin{landscape}

\begin{table*}
  \caption{The spectral indices, the derived brightness temperatures $T_\mathrm{B}$ in Kelvins (K) and luminosity, $L_\mathrm{1.4GHz}$\,(W\,Hz$^{-1}$) across the VLBA, EVN, \textit{e}-MERLIN and VLA. The $\sim5.5\sigma$ sources are italicised and highlighted with an asterik. Where the spectral index was unavailable, the $\alpha_{1.5}^{5.5}$ median value of the sample, $-0.57$, was used to derive the luminosity of the source.}
  \centering
  \begin{tabular}{llllllllll}
    \hline
     Source ID  &  $\alpha$ & \multicolumn{4}{c}{$T_\mathrm{B}$ (K)}&\multicolumn{4}{c}{$ L_\mathrm{1.4GHz}$\,(W\,Hz$^{-1}$)}\\
    
         &  &  VLBA & EVN & \textit{e-}MERLIN & VLA &  VLBA & EVN & \textit{e-}MERLIN & VLA\\ 
        \hline
J123751.2+621919& -- &	$>6.7\, \times10^{6} \, $& $\mathrm{Unresolved}$ & 5.3$\times 10^{3}$ &	5.6$\times 10^{2}$ & $7.3\times 10^{24}$ & $>8.9\times 10^{24}$ & $5.3\times 10^{24}$ & $7.6\times 10^{24}$\\

J123746.6+621738& -- &	4.8$\times 10^{7}$ & $\mathrm{No\,data}$ & 1.1$\times 10^{5}$&	6.3$\times 10^{3}$& $1.4\times 10^{26}$& $\mathrm{No\,data}$  & $1.3\times 10^{26}$ &  $2.3\times 10^{26}$ \\

J123734.5+621723$^\ast$& -- & $\mathrm{Unresolved}$ 	& $\mathrm{Undetected}$ & 2.4$\times 10^{2}$ &	8.6$\times 10^{1}$ & $1.2\times 10^{24}$& $\mathrm{Undetected}$  & $5.9\times 10^{23}$& $1.3\times 10^{24}$\\

J123721.2+621129& 0.04 & 3.6$\times 10^{7}$ &	$>9.0\times 10^{7}$ &	5.2$\times 10^{4}$ &	2.8$\times 10^{3}$& $2.4\times 10^{25}$ & $3.4\times 10^{25}$ & $2.8\times 10^{25}$  & $3.4\times 10^{25}$\\

J123716.7+621733& -0.84 &			5.0$\times 10^{6}$  &	7.0$\times 10^{6}$ &	3.9$\times 10^{4}$ &	1.3$\times 10^{3}$ & $1.1\times 10^{25}$& $1.7\times 10^{25}$ & $1.5\times 10^{25}$ & $2.1\times 10^{25}$ \\

J123716.4+621512& -0.18	&	4.8$\times 10^{6}$	& 2.0$\times 10^{6}$	& 1.3$\times 10^{4}$& 5.4$\times 10^{2}$ & $9.3\times 10^{23}$ & $1.1\times 10^{24}$ & $1.2\times 10^{24}$ & $1.5\times 10^{24}$ \\

J123713.8+621826& -0.64 &  4.6 $\times 10^{7}$ &	$>2.0\times 10^{8}$ &	1.6$\times 10^{5}$ &	6.2$\times 10^{3}$ & $4.7\times 10^{26}$ & $6.7\times 10^{26}$ & $4.9\times 10^{26}$ & $6.6\times 10^{26}$ \\

J123701.1+622109&  -- &	1.8$\times 10^{7}$	 & $>1.9\times 10^{6}$ & $\mathrm{Undetected}$ &		1.1$\times 10^{3}$ & $5.1\times 10^{24}$ & $3.6\times 10^{24}$ & $\mathrm{Undetected}$ & $8.4\times 10^{24}$ \\

J123700.2+620909& -0.84	&	8.1$\times 10^{6}$ &	5.0$\times 10^{6}$ &	7.5$\times 10^{4}$ &	3.0$\times 10^{4}$ & $7.5\times 10^{25}$ & $7.1\times 10^{25}$ & $1.0\times 10^{26}$ & $1.3\times 10^{26}$ \\

J123659.3+621832& -1.18 &	1.5$\times 10^{8}$ &	$>1.0\times 10^{9}$ &	1.5$\times 10^{6}$ &	9.3$\times 10^{2}$ & $1.4\times 10^{27}$ & $1.9\times 10^{27}$ & $2.7\times 10^{27}$ & $2.0\times 10^{27}$ \\

J123652.9+621444& -0.10 &   $\mathrm{Unresolved}$ &	2.0$\times 10^{6}$ &	1.0$\times 10^{4}$ &	9.1$\times 10^{2}$ & $2.0\times 10^{23}$& $3.1\times 10^{23}$ & $3.6\times 10^{23}$ & $5.5\times 10^{23}$\\

J123646.3+621404& -0.39 &	8.2$\times 10^{6}$	& $>2.0\times 10^{7}$ &	4.7$\times 10^{4}$ &	2.7$\times 10^{3}$ &$4.5\times 10^{24}$ &  $6.0\times 10^{24}$ & $6.7\times 10^{24}$ & $8.5\times 10^{24}$ \\

J123644.4+621133& -0.59 &	 $\mathrm{Unresolved}$ 	& $>1.0\times 10^{8}$ &	4.5$\times 10^{4}$ &	1.5$\times 10^{3}$ & $1.0\times 10^{25}$ & $1.7\times 10^{25}$ & $2.5\times 10^{25}$ & $3.4\times 10^{25}$\\

J123642.1+621331& -1.02 &		4.3$\times 10^{6}$	& 3.0$\times 10^{6}$ &	7.6$\times 10^{4}$ &	3.1$\times 10^{3}$ & $4.4\times 10^{26}$ & $4.7\times 10^{26}$ & $6.3\times 10^{26}$ & $9.3\times 10^{26}$ \\

J123640.5+621833& -1.00 &		3.8$\times 10^{6}$	& 3.0$\times 10^{6}$	& 2.5$\times 10^{4}$ &	3.2$\times 10^{2}$ & $1.4\times 10^{25}$ & $1.0\times 10^{25}$ & $1.5\times 10^{25}$  & $2.3\times 10^{25}$ \\

J123634.5+621241$^\ast$& -0.89 &  >8.6$\times 10^{6}$ 	& $\mathrm{Undetected}$&  7.9$\times 10^{2}$ &	1.1$\times 10^{2}$ & $6.7\times 10^{24}$ & $\mathrm{Undetected}$ & $2.1\times 10^{25}$ & $1.4\times 10^{25}$\\

J123627.2+620605$^\ast$& -- & $\mathrm{Unresolved}$ & $\mathrm{Undetected}$  & $\mathrm{Unresolved}$ & $\mathrm{No\,Redshift}$& $\mathrm{No\,Redshift}$& $\mathrm{Undetected}$& $\mathrm{No\,Redshift}$ & $\mathrm{No\,Redshift}$ \\

J123624.6+620728$^\ast$& -- & $2.7\times 10^{6}$	 &$\mathrm{Undetected}$ & $\mathrm{Unresolved}$ &$\mathrm{Undetected}$ & $7.2\times 10^{23}$ &$\mathrm{Undetected}$ & $1.5\times 10^{23}$ & $\mathrm{Undetected}$ \\	

J123623.5+621642& -0.54 &		1.5$\times 10^{7}$	& 2.0$\times 10^{7}$ &	9.7$\times 10^{4}$ &	2.3$\times 10^{3}$ & $4.0\times 10^{25}$ & $6.1\times 10^{25}$ & $4.9\times 10^{25}$ & $6.5\times 10^{25}$\\

J123622.5+620653& -0.004 &	3.4$\times 10^{7}$	& $2.0\times10^{6}$ &	3.6$\times 10^{4}$ &	2.0$\times 10^{3}$ & $1.7\times 10^{25}$ & $1.3\times 10^{25}$ & $1.6\times 10^{25}$ & $2.2\times 10^{25}$ \\

J123620.2+620844&  -0.12 &		6.5$\times 10^{6}$&	$>2.0\times 10^{7}$& $>4.1\times 10^{4}$  &		9.6$\times 10^{2}$ & $3.0\times 10^{24}$ & $5.4\times 10^{24}$& $13.1\times 10^{24}$ & $4.3\times 10^{24}$\\

J123617.5+621540& -0.52 &		1.7$\times 10^{7}$	& $>3.0\times 10^{7}$ &	1.2$\times 10^{5}$ & 1.3$\times 10^{3}$ & $3.5\times 10^{25}$ & $3.2\times 10^{25}$& $3.0\times 10^{25}$ & $4.0\times 10^{25}$ \\

J123608.1+621035 & -0.39 &	3.4$\times 10^{6}$& 	1.0$\times 10^{6}$	& $>2.9\times 10^{4}$ &	5.4$\times 10^{2}$ & $2.0\times 10^{24}$ & $2.1\times 10^{24}$ & $2.0\times 10^{24}$ & $3.3\times 10^{24}$\\

J123603.2+621110$^\ast$& -0.73 & $\mathrm{Unresolved}$  & $\mathrm{Undetected}$ & 5.5$\times 10^{2}$ &	2.4$\times 10^{2}$ & $8.3 \times 10^{23}$ & $\mathrm{Undetected}$ & $1.8\times 10^{24}$ & $2.0\times 10^{24}$\\

     \hline
  \end{tabular}
  \label{Tab: Tb} 
 \end{table*}
\end{landscape}

\begin{figure*}
\centering
\includegraphics[width=1\linewidth, height=24cm, keepaspectratio=true]{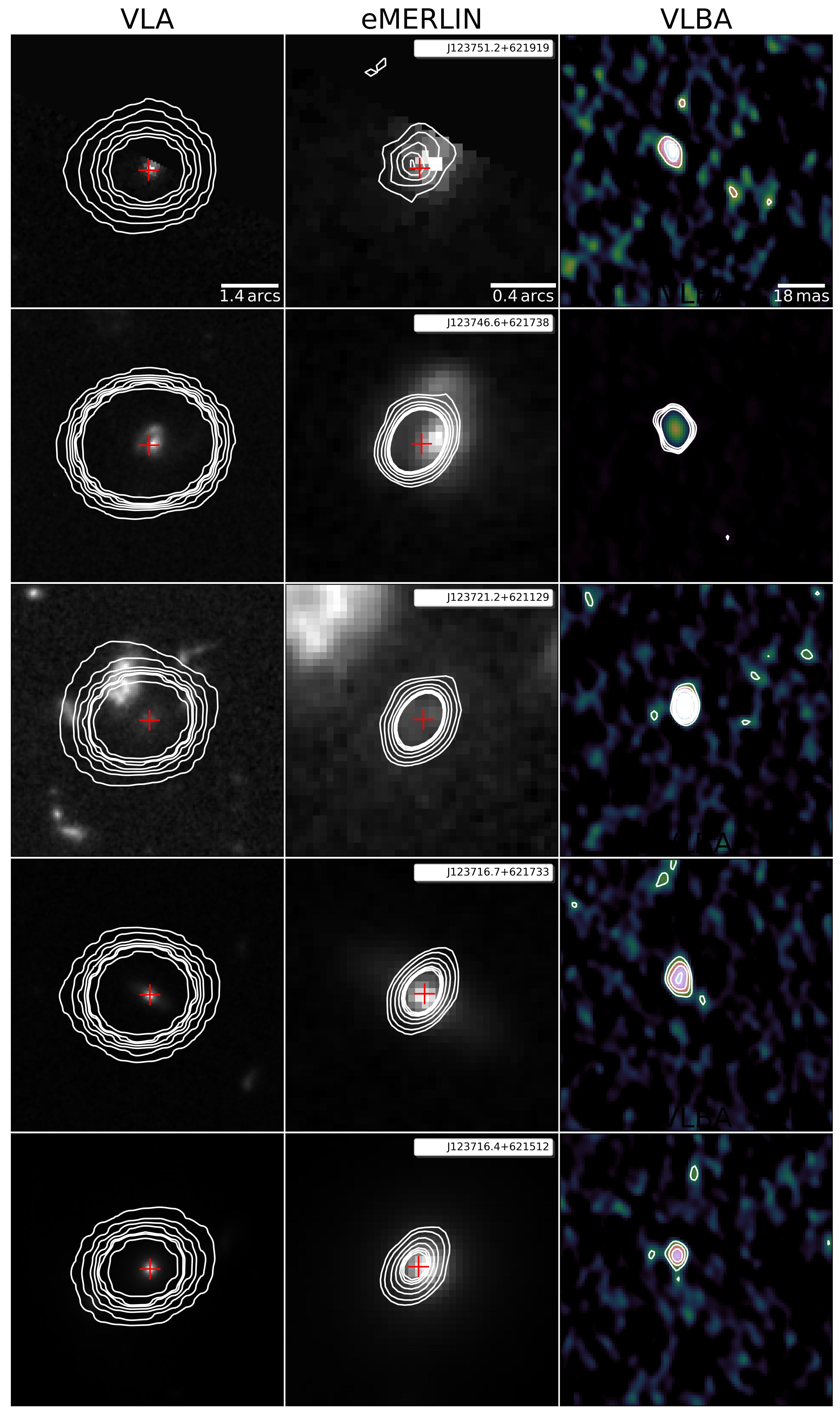}
\end{figure*}

\begin{figure*}
\centering
\includegraphics[width=1\linewidth, height=24cm, keepaspectratio=true]{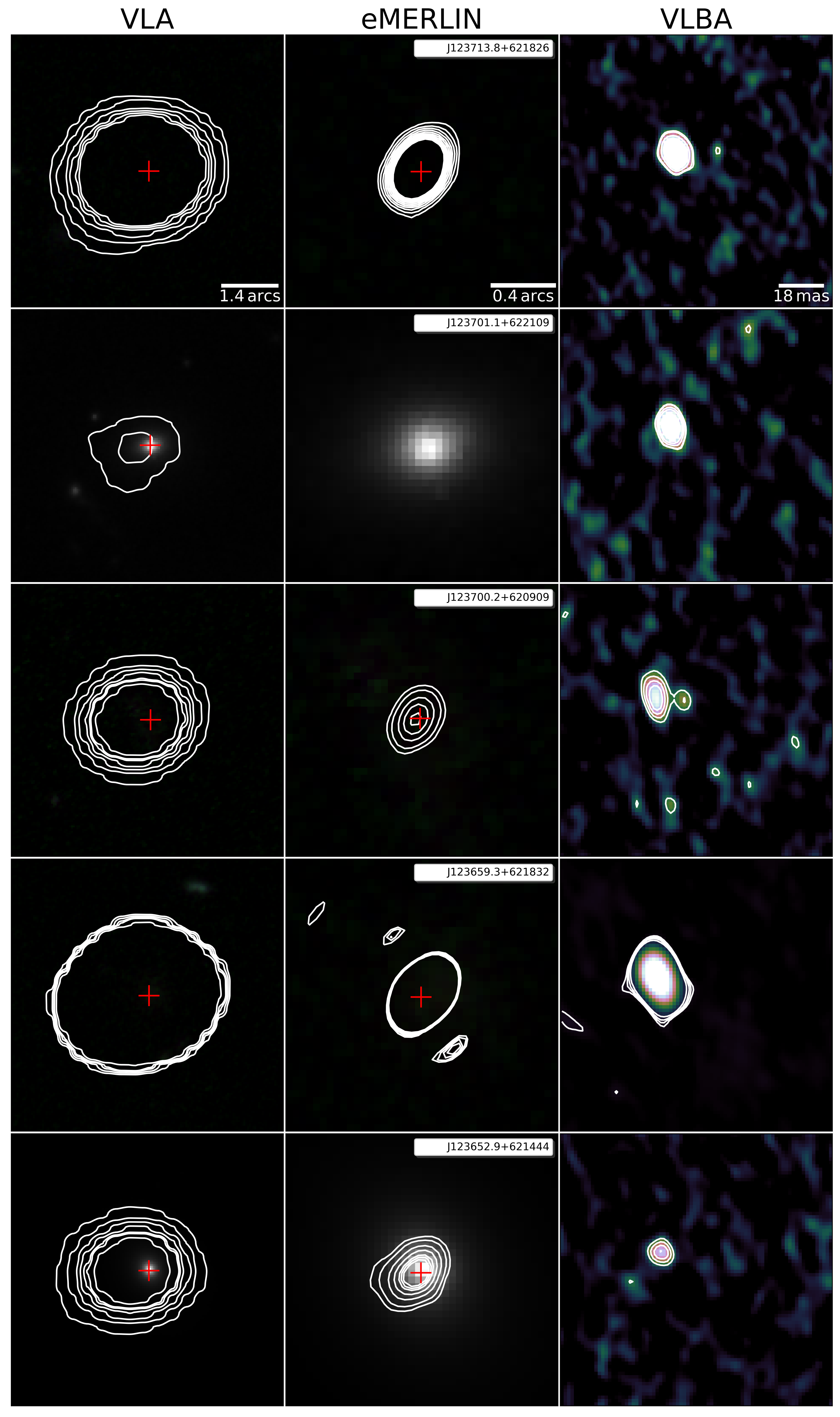}
\end{figure*}

\begin{figure*}
\centering
\includegraphics[width=1\linewidth, height=24cm, keepaspectratio=true]{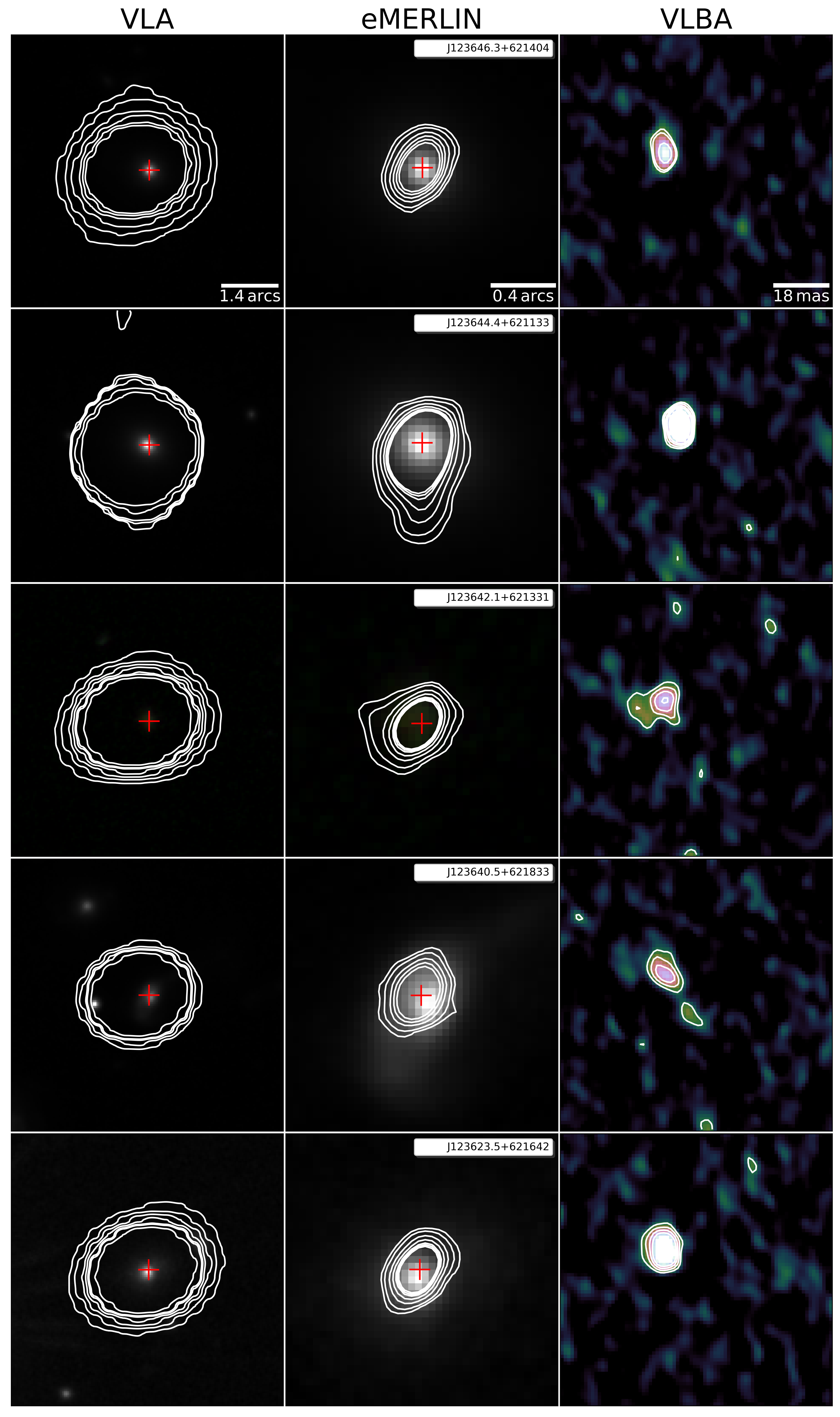}
\end{figure*}

\bsp	
\label{lastpage}
\end{document}